\documentclass[preprint,12pt]{elsarticle}

\usepackage{amssymb}
\usepackage{amsmath}
\usepackage{xspace}
\usepackage{url}
\usepackage{color}
\usepackage[most]{tcolorbox}
\definecolor{darkred}{rgb}{0.60,0,0}
\definecolor{darkgreen}{rgb}{0,0.44,0}
\definecolor{darkblue}{rgb}{0,0,0.60}
\definecolor{grey}{rgb}{0.8,0.8,0.8}
\newcommand{\tcb}[1]{\colorbox{white}{\parbox{4ex}{\textcolor{darkblue}{\small #1}}}}
\newcommand{\tcr}[1]{\colorbox{white}{\parbox{4ex}{\textcolor{darkred}{\small #1}}}}
\newcommand{\tck}[1]{\colorbox{grey}{\parbox{4ex}{\textcolor{black}{\small #1}}}}
\newcommand{\tckl}[1]{\colorbox{grey}{\parbox{6ex}{\textcolor{black}{\small #1}}}}
\newcommand{\ssf}[1]{{\sf \small #1}}
\usepackage{multirow}

\usepackage{algpseudocode}
\usepackage{algorithm}
\usepackage{listings}
\renewcommand{\H}{$\mathcal{H}$\xspace}
\newcommand{\hmat}{\H-matrix\xspace}
\newcommand{\hmats}{\H-matrices\xspace}
\newcommand{\harith}{\H-arithmetic\xspace}

\journal{Journal of Computational Science}

\begin{document}

\begin{frontmatter}



\title{Exploiting Nested Task-Parallelism in\\ the \H-LU Factorization}
       
\author[uji]{Roc\'{\i}o~Carratal\'a-S\'aez}
\ead{rcarrata@uji.es}
\author[kiel]{Sven~Christophersen}
\ead{svc@informatik.uni-kiel.de}
\author[uji]{Jos\'e~I.~Aliaga}
\ead{aliaga@uji.es}
\author[bsc]{Vicen\c{c}~Beltran}
\ead{vbeltran@bsc.es}
\author[kiel]{Steffen~B\"orm}
\ead{sb@informatik.uni-kiel.de}
\author[uji]{Enrique~S.~Quintana-Ort\'{\i}}
\ead{quintana@uji.es}

\address[uji]{Depto. Ingenier\'{\i}a y Ciencia de Computadores,
             Universidad Jaume I, Castell\'on, Spain.}
\address[kiel]{Mathematisches Seminar, Universit\"at zu Kiel, Germany.}
\address[bsc]{Barcelona Supercomputing Center, Spain.}       


\author{}

\address{}

\begin{abstract}
We address the parallelization of the LU factorization of hierarchical matrices (\hmats) arising from boundary element methods.
Our approach exploits task-parallelism via the OmpSs programming model and runtime, which discovers the data-flow 
parallelism intrinsic
to the operation at execution time, 
via the analysis of data dependencies based on the memory addresses of the tasks' operands. This is especially challenging
for \hmats, as the structures containing the data vary in dimension during the execution. We tackle this issue by 
decoupling the data structure from that used to detect dependencies. Furthermore, we leverage the support
for weak operands and early release of dependencies,
recently introduced in OmpSs-2, to accelerate the execution of parallel codes with nested task-parallelism and 
fine-grain tasks.
\end{abstract}

\begin{keyword}
Hierarchical linear algebra \sep
LU factorization \sep
nested task-parallelism \sep
task dependencies \sep
multi-threading \sep
multicore processors \sep
boundary element methods (BEM)



\end{keyword}

\end{frontmatter}


\section{Introduction}

Hierarchical matrices (or \hmats)~\cite{Hackbusch:1999:SMA:303815.303816} provide a useful
mathematical abstraction to tackle
problems arising in boundary element methods,
elliptic partial differential operators, and related integral equations, among others~\cite{Hack09}.
Concretely, for many of these applications, \hmats and the associated \harith methods offer efficient numerical tools
to store an $n \times n$ matrix using only 
$O(n\,k\,\log n)$ elements 
and compute matrix factorizations with a cost of
$O(n\,k^2\,\log^2 n)$ floating-point operations (flops). 
In these cost expressions, $k$ denotes the local rank for
subblocks of the matrix, which can be tuned to trade off accuracy of the approximation for
storage and computational costs~\cite{GrasHack:Arithm}.

The development of linear algebra methods for \hmats has been an active area of research during the past two decades, having produced a wide number of packages
such as HLib, H2Lib and HLibPro as well as a collection of individual 
routines. 
Part of this software directly relies on the kernels from the
{\em Basic Linear Algebra Subprograms} (BLAS)~\cite{BLAS3} to compute fundamental dense linear algebra (LA) operations.
As a result, when linked with a multi-threaded instance of the BLAS,
these \H-LA routines can seamlessly run in parallel on current multicore processors.
However, this solution can extract a limited amount parallelism, bounded to that present in the individual BLAS kernels.
\textcolor{black}{Furthermore, for \H-matrices arising from real applications, low-rank blocks ``dominate'' the data structure.
While this confers \H-matrices/methods their appealing low storage and computational costs, unfortunately, it also limits further the parallelism that can be extracted
from individual kernels. In particular, \H-arithmetic involves memory-bounded kernels which, in general, do not benefit from a multi-threaded
execution. The bottom-line is that, for practical \H-applications, it becomes necessary to exploit parallelism at a higher level.}


In the last years, exploiting task-parallelism 
has been exposed as an appealing coarse-grain approach 
for the solution of dense and sparse linear systems on multi-threaded 
architectures~\cite{Buttari200938,Quintana:2008:PMA,BadiaHLPQQ09,AliBBBQ14,agullo:hal-01333645}.
These approaches discover task parallelism dynamically (at execution time) via a runtime but rely on a sequential implementation of the numerical kernels
(in the dense case, BLAS) to execute the individual operations.
Although similar, the factorization of \hmats for linear systems presents some specific challenges when the aim is to extract 
task-parallelism.
First, the ``recursive nature'' of \hmats makes the detection and efficient exploitation of 
nested task-parallelism a complex endeavor.
\textcolor{black}{Second, handling low-rank matrices requires specialized data structures that can vary (grow/shrink in size), at execution time,
with low overhead. This is particularly important given the low cost of \H-arithmetic algorithms.}

In~\cite{7965168}, we presented two prototype task-parallel versions of the \H-LU factorization 
using the OpenMP and OmpSs programming models~\cite{openmpweb,ompssweb}. Our initial implementations presented several drawbacks, that we
overcome in this work, making the following specific contributions:
\begin{itemize}
\item The prototype implementations in~\cite{7965168} assumed that the blocks of the \hmat were either dense or null. No specialized
      data structures and \harith for low-rank blocks were therefore involved in the factorization. 
      In comparison, in the present work we parallelize the \H-LU factorization
      as implemented in the sequential version of H2Lib, with problems involving low-rank blocks and, therefore, 
      low-rank storage and real \harith.
\item An additional consequence of targeting the \H-LU factorization in H2Lib is the need to accommodate low-rank data structures that can change their dimensions at execution time. 
      This is particularly challenging for a runtime-based parallelization because task dependencies are detected via an analysis of the memory addresses of the tasks'
      operands. To address this problem, we propose the use of an auxiliary ``skeleton'' array, which reflects the block hierarchy of the \hmat, 
      and can be leveraged to identify task dependencies. This additional data structure is built before the execution commences, at low cost, and remains
      unchanged during the complete execution, independently of the modifications on the structures containing the actual data due to the use of \harith.
\item The task-parallel implementation 
      developed in our past work~\cite{7965168}, based on OmpSs, forced us to operate on fine-granularity tasks 
      with operands that were stored in contiguous regions of memory.
      The practical consequence of this constraint is that it was not possible to exploit
      the nested task-parallelism intrinsic to the \H-LU factorization.
      In the present work, we address these problems using the new OmpSs-2 model,
      with explicit support for weak dependencies and early release to take advantage 
      of fine-grained nested parallelism~\cite{7967171}.
\end{itemize}

The rest of the paper is structured as follows. 
In Sections~\ref{sec:H2Lib} and~\ref{sec:HLU}, we briefly review the structure of \hmats in H2Lib
and introduce a high-level algorithm for the LU factorization of an \hmat, respectively.
(A complete review of \hmats and \harith can be found, for example, in~\cite{Hackbusch:1999:SMA:303815.303816,GrasHack:Arithm}.)
In Section~\ref{sec:pdsec} we re-visit the parallelization of the \H-LU factorization
in~\cite{7965168}, exposing the limitations of those prototype codes.
In Section~\ref{sec:nested} we describe the new task-level parallelization
of the hierarchical factorization, offering details on the use of the skeleton structure to detect task dependencies and the exploitation of 
fine-grained nested parallelism via weak dependencies.
Finally, in Section~\ref{sec:experiments} we provide a complete experimental evaluation of the task-parallel codes, and
in Section~\ref{sec:remarks} we close the paper with a few concluding remarks.


\section{Representation of $\mathcal{H}$-Matrices in H2Lib}
\label{sec:H2Lib}

In this section we briefly introduce the structure
of $\mathcal{H}$-matrices and how they are implemented in H2Lib.\footnote{\url{http://www.h2lib.org/}}
This sequential library, written in C, 
offers a state-of-the-art implementation of \hmat techniques, including
sophisticated data structures, support for \harith operations such as 
multiplication, inversion and factorization,
compression schemes for non-local operators, and fast re-compression algorithms.

The main idea behind \hmats relies on finding a partition over a matrix
into blocks which are either small in dimension or \emph{admissible} in the sense
that they can be stored efficiently using \emph{low-rank data structures} instead of dense ones.
Utilizing low-rank matrices is a prerequisite for reducing the storage and computational 
costs down to log-linear functions on the number of elements and flops, respectively.

Before we can find a partition of the matrix into a set of subblocks, we need
to ``organize'' the degrees of freedom (DoFs) into sets, which can be handled
efficiently via a tree-like structure called \emph{clustertree}.
Therefore, we assume that we know the geometric extent of every degree of
freedom that appears in our application. This frequently reduces to the support
of finite element basis functions.
Armed with that information, we can setup an axis-parallel bounding box 
$\mathcal{B}_t$, which contains the union of all extents corresponding to the cluster $t$. 
This box will now be splitted into two parts along some geometrical dimension,
which yields two independent boxes $\mathcal{B}_{t_1},\, \mathcal{B}_{t_2}$. From there,
we can sort each and every DoF into one of these boxes according to their 
position in space.
Next we process both boxes recursively until the number of
DoFs located in a box falls below a prescribed constant, which we denote by
\emph{leafsize} ($C_{lf}$).
In order to handle all these boxes efficiently, we organize these clusters into 
a tree structure expressed by $\operatorname{sons}(t) = \{t_1, t_2\}$.
This procedure is summarized in Algorithm~\ref{alg:clustertree_construction}.

\begin{algorithm}[h]
\caption{Clustertree construction}
\label{alg:clustertree_construction}
\begin{algorithmic}
\Require{Geometric information about the degrees of freedom is stored within
an array \texttt{dofs} of length \texttt{size}.}
\Ensure{A hierarchical partition of the DoFs is returned via the clustertree
\texttt{t}.}
\Procedure{setup\_clustertree}{dofs, size}
\If {size $> C_{lf}$}
	\State $d \gets $ \Call {find\_splitting\_dimension}{dofs, size}
	\State sons $ \gets $ 2
	\State $t \gets $ \Call {new\_cluster}{dofs, size, sons}
	\State $\{\text{dofs1, dofs2, size1, size2}\} \gets $ 
	  \Call {sort\_dofs}{dofs, size, $d$}
	\State $t_1 \gets $ \Call {setup\_clustertree}{dofs1, size1}
	\State $t_2 \gets $ \Call {setup\_clustertree}{dofs2, size2}
	\State $\operatorname{sons}(t) \gets \{t_1, t_2\}$
\Else
	\State $t \gets $ \Call {new\_leaf\_cluster}{dofs, size}
\EndIf
\State \Return $t$
\EndProcedure
\end{algorithmic}
\end{algorithm}

This hierarchical structure is realized with the following C data structure 
\texttt{cluster} within H2Lib:
\vspace*{1ex}
\begin{lstlisting}[language=C,caption=,morekeywords={uint}]
struct cluster {
    uint     size;
    uint     *dofs
    uint     *bbox_min;
    uint     *bbox_max;   
    cluster  *son;
    uint     sons;
}
\end{lstlisting}
Here \texttt{size} is the number of elements associated with this cluster and
\texttt{dofs} is an array referring to the degrees of freedom.
The bounding box $\mathcal{B}_t$ for a cluster $t$ is stored within the
arrays \texttt{bbox\_min} and \texttt{bbox\_max}, respectively.
In addition,
\texttt{son} and \texttt{sons} represent the tree structure
of the clusters.

In order to identify subblocks of the matrix, which can be approximated by
a low rank representation, we need some \emph{admissibility condition}, which ensures that
we can find an approximation to some prescribed accuracy.
For many cases, particularly for tensor-interpolation, the condition
\begin{equation*}
\operatorname{min}\{
   \operatorname{diam}(t),
   \operatorname{diam}(s)\}
 \leq
   \operatorname{dist}(t,s)
\end{equation*}
guarantees this.
Here $\emph{diam}$ and $\emph{dist}$ denote the Euclidean 
diameter of some cluster and the Euclidean distance between two clusters,
respectively.

During the setup of the \hmat, the partitioning is performed recursively as stated in 
Algorithm~\ref{alg:blocktree_construction}, returning a tree-like 
block structure.
For a detailed construction of such block partition, see~\cite{Hackbusch:1999:SMA:303815.303816,GrasHack:Arithm}.

\begin{algorithm}[h]
\caption{Blocktree construction}
\label{alg:blocktree_construction}
\begin{algorithmic}
\Require{row cluster \texttt{t}, column cluster \texttt{s}.}
\Ensure{A blocktree \texttt{b} is returned for the pair \texttt{(t,s)}.}
\Procedure{setup\_blocktree}{t, s}
\If { \Call {admissible}{t, s} }
	\State b $ \gets$ \Call {new\_admissible\_block}{t, s}
\Else
	\If { \Call {sons}{t} $\neq \emptyset ~\land$ \Call {sons}{s} $ \neq \emptyset$ }
		\State b $ \gets$ \Call { new\_partitioned\_block }{t, s}
		\ForAll {t' $\in $ \Call {sons}{t}, s' $\in $ \Call {sons}{s}}
			\State b[t'][s'] $ \gets$ \Call {setup\_blocktree}{t', s'}
		\EndFor
		\State \Return b
	\Else
		\State b $ \gets$ \Call {new\_inadmissible\_block}{t, s}
	\EndIf
\EndIf
\State \Return b
\EndProcedure
\end{algorithmic}
\end{algorithm}

In agreement with the three cases occurring in Algorithm~\ref{alg:blocktree_construction},
the application of 
Algorithm~\ref{alg:blocktree_construction}, at a given level of the recursion, can produce
either a low-rank block (i.e., a new admissible block), a new recursive partitioning (via the same algorithm), or 
a conventional dense (inadmissible) block.
To handle these cases, the C data type representing a \hmat in the H2Lib follows this structure:
\vspace*{1ex}
\begin{lstlisting}[language=C,caption=,morekeywords={uint}]
struct hmatrix {
    cluster  rc, cc;
    rkmatrix r;
    amatrix  f;
    hmatrix  *son;
    uint     rsons, csons;
}
\end{lstlisting}

\noindent
In this data type, \texttt{rc} and \texttt{cc} respectively correspond to the 
row cluster and column cluster of he current matrix block.
Low-rank matrices are stored in the structure \texttt{rkmatrix},
whereas dense matrices are stored in \texttt{amatrix}.
Partitioned matrices are accommodated using an array pointing to the sons. 
The structure also provides the amount of sons per row and column, stored in \texttt{rsons} and \texttt{csons}, respectively.
 

\section{LU Factorization of $\mathcal{H}$-Matrices}
\label{sec:HLU}

\subsection{Algorithm for the \H-LU factorization}

\begin{figure}[ht]
\centering
\includegraphics[width=0.3\columnwidth]{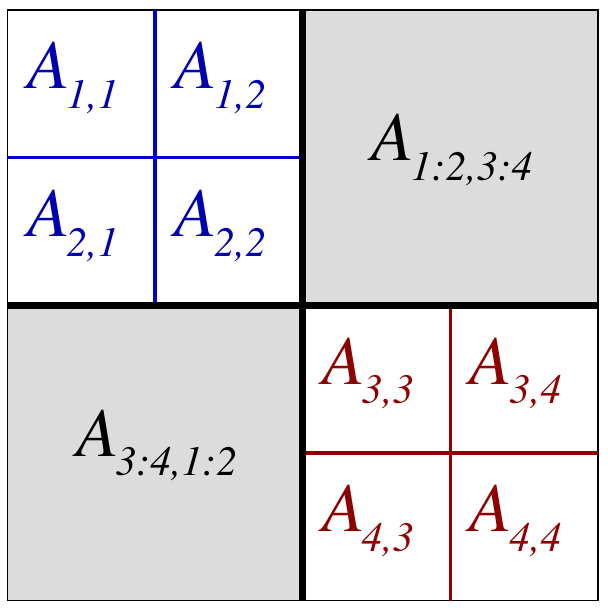}
\caption{$2\times 2$ partitioning of a simple \H-matrix.}
\label{fig:hmat}
\end{figure}

We open this section with a brief review of the algorithm for the \H-LU factorization. 
For this purpose, consider a sample \hmat $A \in \mathbb{R}^{n \times n}$, partitioned as shown
in Figure~\ref{fig:hmat}.  The factorization procedure can be formulated 
as a generalization of the blocked right-looking (RL) algorithm for the LU factorization~\cite{GVL3} 
that exploits the hierarchical structure of the matrix.  

%
In particular, the following sequence of operations
computes the \H-LU factorization of $A$:
\[
\begin{array}{l}
\hline
\mbox{\rm Sequence of operations for the \H-LU factorization of $A$:}\\
\hline
\begin{array}{llcl}
\tcb{\ssf{O1.1}}: & A_{1,1} &=& L_{1,1} U_{1,1}\\
\tcb{\ssf{O1.2}}: & U_{1,2} &:=& L_{1,1}^{-1} A_{1,2}\\ 
\tcb{\ssf{O1.3}}: & L_{2,1} &:=& A_{2,1} U_{1,1}^{-1}\\ 
\tcb{\ssf{O1.4}}: & A_{2,2} &:=& A_{2,2} - L_{2,1} \cdot U_{1,2}\\ 
\tcb{\ssf{O1.5}}: & A_{2,2} &=& L_{2,2} U_{2,2}\\ \hline
\tck{\ssf{O2}}: & U_{1:2,3:4} &:=& L_{1:2,1:2}^{-1} A_{1:2,3:4}\\\hline
\tck{\ssf{O3}}: & L_{3:4,1:2} &:=& A_{3:4,1:2} U_{1:2,1:2}^{-1}\\\hline
\tck{\ssf{O4}}: & A_{3:4,3:4} &:=& A_{3:4,3:4} - L_{3:4,1:2} \cdot U_{1:2,3:4}\\ \hline
\tcr{\ssf{O5.1}}: & A_{3,3} &=& L_{3,3} U_{3,3}\\
\tcr{\ssf{O5.2}}: & U_{3,4} &:=& L_{3,3}^{-1} A_{3,4}\\ 
\tcr{\ssf{O5.3}}: & L_{4,3} &:=& A_{4,3} U_{3,3}^{-1}\\ 
\tcr{\ssf{O5.4}}: & A_{4,4} &:=& A_{4,4} - L_{4,3} \cdot U_{3,4}\\ 
\tcr{\ssf{O5.5}}: & A_{4,4} &=& L_{4,4} U_{4,4}\\ \hline
\end{array}
\end{array}
\]

\paragraph{\bf Dense blocks}
Assuming all blocks are dense, these operations 
correspond to three basic linear algebra building blocks (or computational kernels):
\begin{itemize}
\item LU factorization (e.g., \ssf{O1.1}, \ssf{O1.5} and \ssf{O5.1}); 
\item triangular system solve
      (e.g., with unit lower triangular factor in \ssf{O1.2} and \ssf{O2}; or upper triangular factor in \ssf{O1.3} and \ssf{O3}); and
\item matrix-matrix multiplication
      (\ssf{O1.4}, \ssf{O4}, etc.).
\end{itemize}
In the dense case, the triangular factors $L$ and $U$ overwrite the corresponding entries of $A$ so that, for example,
in \ssf{O2},
the output $U_{1:2,3:4}$ overwrites the input $ A_{1:2,3:4}$.
Furthermore, the diagonal of the unit triangular matrix $L$ only contains ones and it is not explicitly stored.

\paragraph{\bf Low-rank blocks}
In the \H-LU factorization, if any of the matrix blocks is represented in low-rank factorized form, 
the storage will have to be specialized (see section~\ref{sec:H2Lib}) and the operations involving this
block will need to be performed in \harith.

In typical $\mathcal{H}$-matrix implementations, low-rank matrices are
represented in the factorized form $X = A B^*$, where $A$ and $B$ have
only $k$ columns, so the rank of $X$ is bounded by $k$.
In practice, $k$ is significantly smaller than the dimensions of the
original matrix $X$.

The H2Lib packages use the following data type to store low-rank matrices:
\begin{lstlisting}
  typedef struct rkmatrix {
    uint k;        /* Maximal rank */
    amatrix A;     /* Left factor A */
    amatrix B;     /* Right factor B */
  }
\end{lstlisting}

Basic operations for low-rank matrices $X\in\mathbb{R}^{n\times m}$
include:
\begin{itemize}
  \item matrix-vector multiplication $y := X z$,
    performed using $y := X z = A (B^* z)$;
  \item multiplication of $X$ by an arbitrary matrix
    $Z\in\mathbb{R}^{\ell\times n}$, using $Y:=Z X = (Z A) B^*$,
  \item triangular system solve $L Y = X$ or $Y L = X$;
    by applying forward or backward substitution to the $k$ columns
    of $A$ or the $k$ rows of $B$, respectively.
\end{itemize}
Adding two low-rank matrices poses a challenge, since the sum
of two matrices $X_1=A_1 B_1^*$ and $X_2=A_2 B_2^*$, of ranks $k_1$ and $k_2$,
may have a rank of $k_1+k_2$.
Fortunately, in typical applications a low-rank approximation can be
constructed by computing, e.g., a singular value decomposition of
\begin{displaymath}
  X_1 + X_2 = A_1 B_1^* + A_2 B_2^*
  = \begin{pmatrix}
      A_1 & A_2
    \end{pmatrix}
    \begin{pmatrix}
      B_1 & B_2
    \end{pmatrix}^*
\end{displaymath}
and discarding small singular values.
The same approach can be employed to convert an arbitrary matrix
into a factorized low-rank matrix.

From the point of view of peak floating-point performance,
working with factorized low-rank matrices poses a challenge.
Concretely, while the multiplication of two $n\times n$-matrices requires
$2 n^3$ operations, i.e., $n$ operations for each coefficient
transferred from main memory, only $2 k n^2$ operations are required
if one of the factors is a factorized low-rank matrix, i.e., only
$k$ operations for each coefficient.
Therefore we have to deal with the fact that the speed of operations
involving low-rank matrices and, in consequence, $\mathcal{H}$-matrices,
is generally limited by the memory bandwidth, instead of by the floating-point
throughput.



\subsection{Nested dependencies in the factorization}

\begin{figure}[ht]
\centering
\includegraphics[scale=0.7]{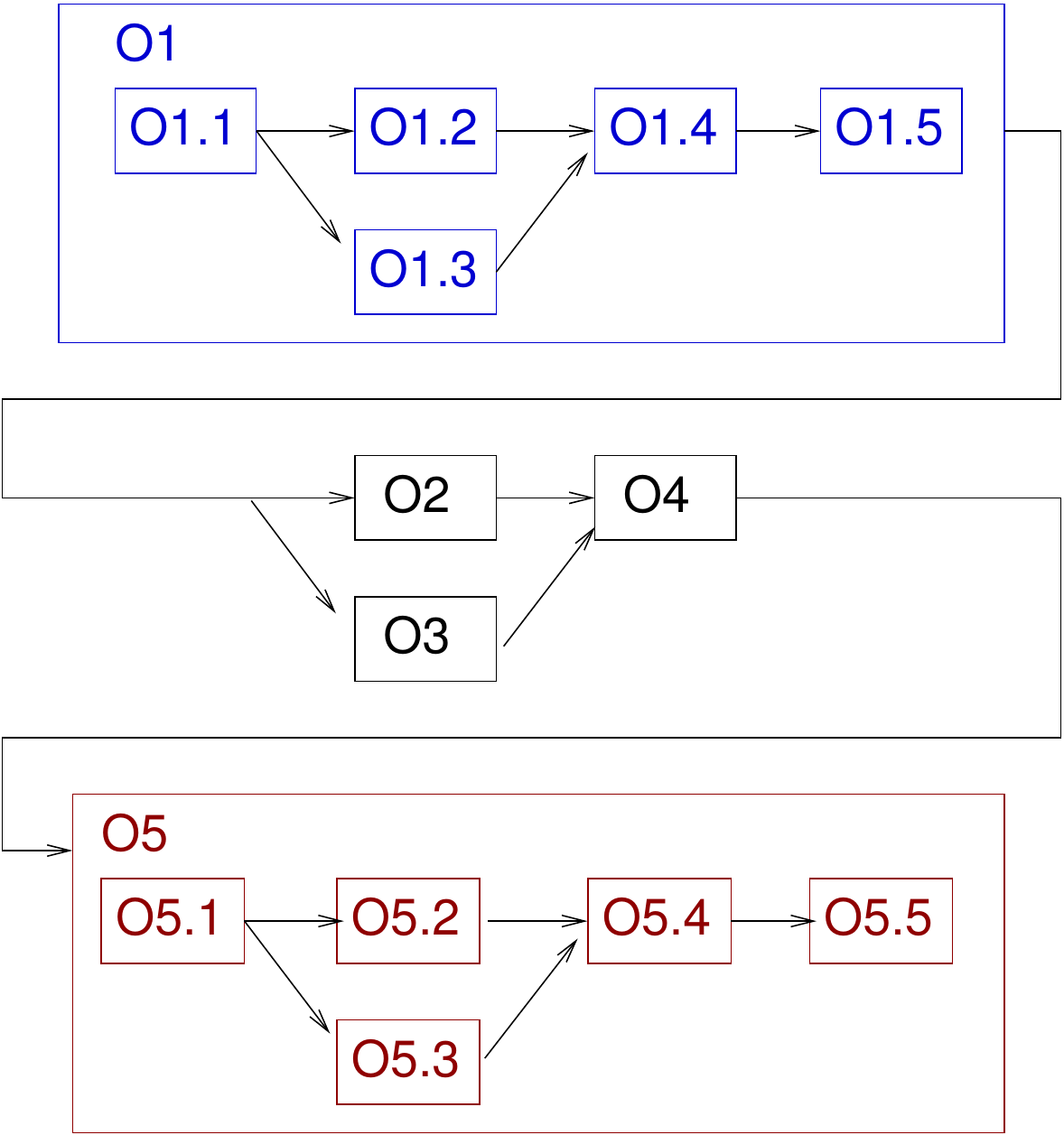}
\caption{Data dependencies in the blocked RL algorithm for the \H-LU factorization.}
\label{fig:dependencies}
\end{figure}

Figure~\ref{fig:dependencies} provides a graphical representation of the 
dependencies among the operations in the \H-LU factorization of the sample matrix $A$,   
exposing the recursion implicit in the operation.
Concretely, the factorization can be initially decomposed
into 5~tasks:
\ssf{O1}, 
\ssf{O2},
\ssf{O3},
\ssf{O4} and
\ssf{O5}, with the dependencies among them displayed in the figure.
The first and last (macro-)tasks, 
\ssf{O1} and
\ssf{O5},
corresponding to the factorizations of the diagonal blocks of $A$, 
can themselves be decomposed into 5~(sub-)tasks each, and reproduce the same dependency pattern as that of the initial
factorization.


We close this section by noting that the presence of low-rank blocks does not affect the dependencies, 
only the particular implementation of the internal operations.
\textcolor{black}{Also, the sample matrix employed in this section  
was specifically chosen to be simple yet useful enough to expose the existence of nested parallelism in
the \H-LU factorization (and discuss how to tackle it in the following sections).
The dependency graph in Figure~\ref{fig:dependencies} seems to show that there is little task-parallelism
to be exploited as, for this particular example, we can only run in parallel 
\ssf{O1.2} with \ssf{O1.3};
\ssf{O2} with \ssf{O3}; and
\ssf{O5.2} with \ssf{O5.3}.
However, this is a direct consequence of the simplicity of the selected example. In contrast, an \H-matrix/block
with a division (partitioning) into $b \times b$ subblocks yields a rapid
explosion of the degree of task-level parallelism that is cubic in $b$, featuring a richer set of dependencies.}


\section{Leveraging Task-Parallelism in OmpSs}
\label{sec:pdsec}

Section~\ref{sec:HLU} and Figure~\ref{fig:dependencies} expose the recursive character of the \H-LU factorization and
the presence of nested task-parallelism. 
%
%
%
In particular, coming back to our sample \H-LU factorization in the
previous section, a natural approach to exploit nested task-parallelism is to annotate
\ssf{O1},
\ssf{O2},
\ssf{O3},
\ssf{O4},
\ssf{O5} each as a task using the OpenMP/OmpSs {\tt task} construct~\cite{openmpweb,ompssweb}. 
Inside
\ssf{O1}, we can then annotate
\ssf{O1.1},
\ssf{O1.2},\ldots,\ssf{O1.5} each as a task; and a similar argument applies
to the decomposition of \ssf{O5} into
\ssf{O5.1},
\ssf{O5.2},\ldots,\ssf{O5.5}.

This section briefly reviews 
the drawbacks of our prototype task-parallel \H-LU implementation 
in~\cite{7965168}
due to the limited support for {\em nested} task-parallelism and hierarchical data structures in OpenMP (version 4.5) and OmpSs (version 16.06).

\subsection{Using representants}

The OpenMP and OmpSs runtimes identify task dependencies, at runtime, via the analysis of the memory addresses of 
the task operands (variables) and their directionality.
In order to specify the dependencies between tasks, in dense linear algebra operations
we can often use a ``representant'' for each task operand, which is then passed to the runtime system
in order to detect these dependencies~\cite{BadiaHLPQQ09}. (This representant is the memory address
of the matrix block computed by the corresponding operation; that is, the top-left entry of the output matrix block.)
We next discuss the problem with this approach in the context of hierarchical matrices.

Let us consider, for example, the dependency
\ssf{O1.1}$\rightarrow$\ssf{O1.2},
between the LU factorization
\[
\begin{array}{rrclrlr}
\mbox{\ssf{O1.1}}: & A_{1,1} &=& L_{1,1} U_{1,1},
\end{array}
\]
and the triangular system solve 
\[
\begin{array}{rrclrlr}
\mbox{\ssf{O1.2}}: & U_{1,2} &:=& L_{1,1}^{-1} A_{1,2};
\end{array}
\]
and the dependency 
\ssf{O1}$\rightarrow$\ssf{O2},
between the LU factorization
\[
\begin{array}{rrclrlr}
\mbox{\ssf{O1}}: & A_{1:2,1:2} &=& L_{1:2,1:2} U_{1:2,1:2},
\end{array}
\]
and the triangular system solve 
\[
\begin{array}{rrclrlr}
\mbox{\ssf{O2}}: & U_{1:2,3:4} &:=& L_{1:2,1:2}^{-1} A_{1:2,3:4}.
\end{array}
\]
(We note here that the composition of the operations
\ssf{O1.1}--\ssf{O1.5} yields the LU factorization
of $A_{1:2,1:2}$ specified in 
\ssf{O1}.)

For simplicity, let us assume that all the blocks involved in these operations 
are dense. (The analysis of the dependencies for low-rank blocks is analogous.)
The problem with the use of representants is that 
it is not possible to distinguish a dependency with
input $A_{1:2,1:2}$ from one that has its origin in the input $A_{1,1}$.
In particular, since both $A_{1:2,1:2}$  and $A_{1,1}$  share the same representant,
with this technique it is not possible to know whether 
\ssf{O1.2} and 
\ssf{O2} 
depend either on 
\ssf{O1.1} or
\ssf{O1}.

\subsection{Leveraging regions}

OmpSs
offers flexibility to specify the shapes/dimensions of the input/output operands passed to a task as
{\em regions}, which can then be used to detect dependencies between the tasks. 
In principle, it might seem that
this mechanism could be leveraged to avoid the ambiguity due to the use of representants.
However, 
the following discussion illustrates that this is still insufficient for H2Lib.

To expose the problem, consider again the dependencies
\ssf{O1.1}$\rightarrow$\ssf{O1.2} and
\ssf{O1}$\rightarrow$\ssf{O2}
where, for simplicity, we still assume that all blocks involved in these operations
are dense.
To tackle this case, it might seem that we could simply specify the dimensions of the operands.
For example, in OmpSs, 
the lower triangular system solves 
\ssf{O1.2} and \ssf{O2}
could be annotated as
\vspace*{1ex}
\begin{lstlisting}[language=C,caption=,morekeywords={pragma,omp,task}]
#pragma omp task in( L[0;M*M] ), inout( B[0;M*P] ) 
void task_ltrsm( int M, int P, double *L, int LDL, 
                               double *B, int LDB )
\end{lstlisting}
where {\tt L} is (the memory address of) the {\tt M}$\times${\tt M}
lower triangular factor and 
{\tt B} is (the memory address of) the {\tt M}$\times${\tt P} right-hand side.


The problem with this solution is that,
in H2Lib, the entries of a (dense) block which is further partitioned into subblocks 
(as is the case for $A_{1:2,1:2}$) are not stored contiguously in memory. Therefore, the use of a region to specify
the memory address of the contents of such block is useless. 
The same problem appears for partitioned low-rank blocks.

Our workaround to this problem in~\cite{7965168} was to divide 
the triangular system solve in \ssf{O2} into four tasks, each updating one of the four blocks
of $U_{1:2,3:4}$.
Unfortunately, this solution 
implies the need to explicitly decompose all tasks in the \H-LU factorization to operate with blocks of the ``base'' granularity, 
so that a region only spans 
data which is contiguous in memory.
The practical consequence is that, with that approach, it was not truly possible to exploit nested task-parallelism.
Furthermore, in case of small leaf blocks, the overhead introduced by the dependency-detection mechanism can be considerable, 
reducing the performance of the solution.


\section{Extended Support for Nested Task-Parallelism in OmpSs-2}
\label{sec:nested}

\subsection{Dealing with non-contiguous regions}

The (end of the) previous section identified a major problem when tackling
nested parallelism and hierarchical data structures which do not lie contiguously in memory. 
This issue is difficult to address, as it is rooted on the 
hierarchical nature of the problem and the use of \H-arithmetic, which derives in the need to embody a data structure 
that can vary at runtime. With these premises, it becomes necessary 
to maintain a tree-like structure of the matrix contents (see Section~\ref{sec:HLU}), where only 
the ``leaf'' blocks (either dense or low-rank) store their data 
contiguously in memory. As a result, we cannot leverage this data structure to specify dependencies
between tasks involving non-leaf blocks.

Our solution to this problem is application-specific 
(but can be leveraged in scenarios involving dynamic and/or complex data 
structures~\cite{AliBBBQ14}) 
and consists of an auxiliary skeleton data structure that 
reflects the block structure of the \hmat. In particular, this data structure can be realized using
an array with one representant per leaf (i.e., non-partitioned) block in the original matrix,
where the representants that pertain to the same block appear in contiguous positions of memory. 
For the particular simple example in Figure~\ref{fig:hmat} this means that, in order to detect
dependencies, we use an additional array with representants for 
\[
\fcolorbox{darkblue}{white}{\tcb{$A_{1,1}$}}
\fcolorbox{darkblue}{white}{\tcb{$A_{1,2}$}}
\fcolorbox{darkblue}{white}{\tcb{$A_{2,1}$}}
\fcolorbox{darkblue}{white}{\tcb{$A_{2,2}$}}
\fcolorbox{black}{grey}{\tckl{$A_{3:4,1:2}$}} 
\fcolorbox{black}{grey}{\tckl{$A_{1:2,3:4}$}}
\fcolorbox{darkred}{white}{\tcr{$A_{3,3}$}}
\fcolorbox{darkred}{white}{\tcr{$A_{3,4}$}}
\fcolorbox{darkred}{white}{\tcr{$A_{4,3}$}}
\fcolorbox{darkred}{white}{\tcr{$A_{4,4}$}}
\]
appearing in that specific order.
Operating in this manner, we decouple the mechanism to detect the dependencies (based on the previous array) from
the actual layout of the data in memory, which can vary during the execution.

With this solution, the ambiguity between 
\ssf{O1.1} and
\ssf{O1} when dealing with the dependencies
\ssf{O1.1}$\rightarrow$\ssf{O1.2} and
\ssf{O1}$\rightarrow$\ssf{O2} is easily tackled.
Concretely, although both operands share the same base address in memory
(that of $A_{1,1}$ in the skeleton array), the region for
\ssf{O1.1} comprises a single representant while that of
\ssf{O1} comprises four representants in the skeleton array.
We emphasize that these representants are stored contiguously in memory and this skeleton data structure does not
vary during the execution (in contrast with the structure storing the actual data).
Therefore, it can be built before the operations commence, and the cost of assembling it can be amortized over enough
computation.

\subsection{Weak dependencies and early release}

The tasking model of OpenMP 4.5 supports both nesting and the definition of dependences between sibling tasks. Many operations with 
\hmats are recursive, so the natural strategy to parallelize them is to leverage task nesting. However, this top-down approach has some drawbacks since combining nesting with dependencies usually requires additional measures to enforce the correct coordination of dependencies across nesting levels. For instance, most non-leaf tasks need to include a 
{\tt taskwait} construct 
at the end of their code. While these measures enforce the correct order of execution, as a side effect, they also constrain both the generation and discovery of task parallelism.
In this paper we leverage the enhanced tasking model recently implemented in OmpSs-2~\cite{7967171} to exploit both nesting and fine-grained data-flow parallelism.

The OmpSs-2 tasking model introduces two major features: {\it weak dependencies} and {\it early release} of dependencies.
The dependencies due to task operands annotated as {\rm weak} are ignored by the runtime when determining whether a task is ready to be executed. 
This is possible because operands marked as {\rm weak} can only be read or written by child tasks. 
Using {\rm weak} dependencies, subtasks can be thus instantiated earlier and in parallel.
The {\rm early} release of dependencies allows a fine-grained release of dependencies to sibling tasks. 
Concretely, with this advanced release, when a task ends, it immediately releases the dependencies that are not currently used by any of its child tasks. 
Furthermore, as soon as child tasks finish, they release the dependencies that are not currently used by any of their sibling tasks. 

To further clarify this, we remark that
the correct use of task nesting and dependencies has to obey the following rule to avoid data-races between  tasks that are second (or above)-degree relative: the dependency set of a child task has to be a subset of the dependency set of its parent task. Only dependencies declared on data that is not available in the scope of the parent task, such as data dynamically allocated when the body of the parent task is executed, are excluded from this rule. Although this rule guarantees the correctness of the execution, it usually introduces artificial coarse-grained dependencies between sibling tasks, which are only required to enforce the proper synchronization of their sibling tasks.

To address the previous issue, we can leverage weak dependencies because this type of dependencies are just ignored by the runtime when determining whether a task is ready to be executed. This is possible because operands marked as weak can only be read or written by child tasks. Using weak dependencies, more tasks can be thus instantiated earlier and in parallel,
and we can avoid the 
insertion of a {\tt taskwait} construct, at the end of each parent task, to enforce a barrier which synchronizes all the child tasks before releasing all the dependencies.

By combining these two contributions, dependencies can cross the boundaries initially set up by the nesting contexts. The resulting behavior is equivalent to performing all the dependency analysis on a single domain. 
Achieving a similar effect in OmpSs eliminated the possibility of nesting. 
In addition, that approach also reduced the programmability and restricted 
the instantiation of tasks to a single generator. 
In constrast, the dependency model of OmpSs-2 can extract the same amount of task parallelism, without impairing programmability and 
without the loss of the parallel generation of work that is possible through nesting.

\begin{figure}[ht]
\centering
\includegraphics[width=0.3\columnwidth]{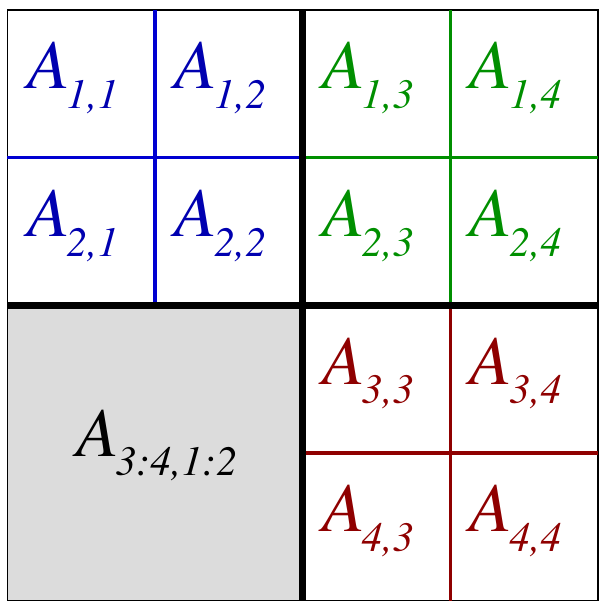}
\caption{Alternative $2\times 2$ partitioning of a simple \H-matrix.}
\label{fig:hmat2}
\end{figure}

In order to discuss the implications of these two advanced features of OmpSs-2 on the \H-LU factorization, let us consider 
the re-partitioning of the initial sample matrix $A$ as shown in Figure~\ref{fig:hmat2}.
(Compared with the initial case in Figure~\ref{fig:hmat}, an additional $2 \times 2$ partitioning has been 
imposed here on the top-right block $A_{1:2,3:4}$.)
Correspondingly, the initial update (see the sequence of operations for the \H-LU factorization in
Section~\ref{sec:HLU})
\[
\begin{array}{rrcl}
\mbox{\ssf{O2}}: & U_{1:2,3:4} &:=& L_{1:2,1:2}^{-1} A_{1:2,3:4}
\end{array}
\]
can be further decomposed into the six suboperations:
\[
\begin{array}{rrcl}
\mbox{\ssf{O2.1}}: & U_{1,3} &:=& L_{1,1}^{-1} A_{1,3}, \\ [0.08in]
\mbox{\ssf{O2.2}}: & U_{1,4} &:=& L_{1,1}^{-1} A_{1,4}, \\ [0.08in]
\mbox{\ssf{O2.3}}: & A_{2,3} &:=& A_{2,3} - L_{21} \cdot U_{1,3}, \\[0.08in]
\mbox{\ssf{O2.4}}: & A_{2,4} &:=& A_{2,4} - L_{21} \cdot U_{1,4} \\[0.08in]
\mbox{\ssf{O2.5}}: & U_{2,3} &:=& L_{2,2}^{-1} A_{2,3}, \quad \mbox{\rm and} \\[0.08in]
\mbox{\ssf{O2.6}}: & U_{2,4} &:=& L_{2,2}^{-1} A_{2,4}.
\end{array}
\]

In the application of nested parallelism to this scenario, we again assume that
\ssf{O1},
\ssf{O2},
\ssf{O3},
\ssf{O4},
\ssf{O5} are each annotated as a (coarse-grain) task, and the decompositions
of 
\ssf{O1},
\ssf{O2},
\ssf{O5}
respectively produce the operations in
\ssf{O1.1}--\ssf{O1.5},
\ssf{O2.1}--\ssf{O2.6},
\ssf{O5.1}--\ssf{O5.5}, each annotated as a (fine-grain) task.

A rapid analysis of the new scenario reveals that, for example, the coarse-grain dependency
\ssf{O1}$\rightarrow$\ssf{O2}
boils down (among others) to the finer-grain cases
\ssf{O1.1}$\rightarrow$\{\ssf{O2.1}, \ssf{O2.2}\}, as 
the former operation (LU factorization)
\[
\begin{array}{rrcl}
\mbox{\ssf{O1.1}}: & A_{1,1} &:=& L_{1,1} U_{1,1}
\end{array}
\]
yields the unit lower triangular factor
$L_{1,1}$ required by the latter two
operations (triangular solves)
\ssf{O2.1}, \ssf{O2.2}.

\begin{figure}[ht]
\centering
\includegraphics[scale=0.7]{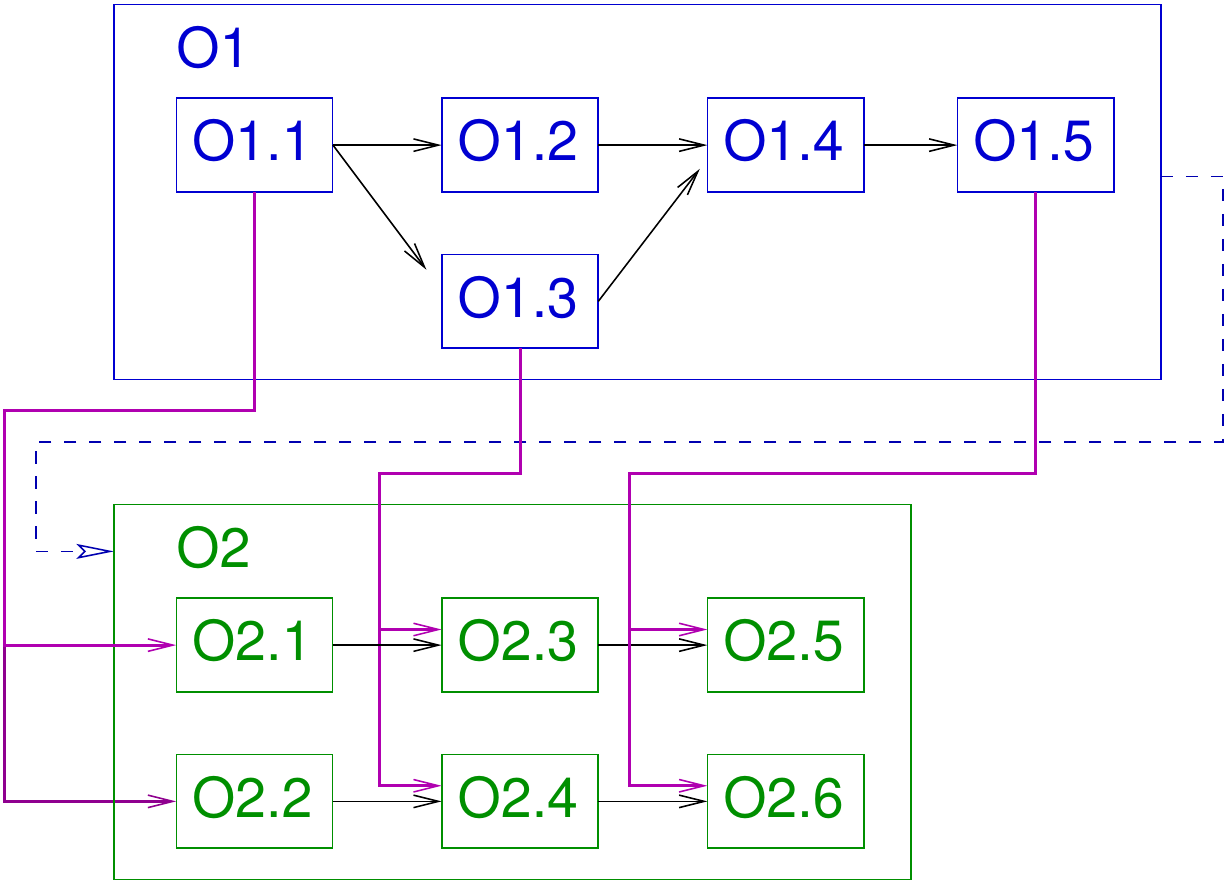}
\caption{Data dependencies between tasks \ssf{O1} and \ssf{O2} of the blocked RL algorithm for the \H-LU factorization. The black solid lines specify ``internal'' strong
         dependencies; the pink solid lines, strong dependencies crossing task boundaries; and the blue dashed line, the weak dependency.}
\label{fig:dependencies_2}
\end{figure}

The problem with OmpSs and OpenMP~4.5 is that ensuring a correct result requires the introduction of a {\tt taskwait}
at the end of the code of 
\ssf{O1} in order to guarantee a correct result.
In contrast, the support for weak dependencies and early release in OmpSs-2 implies that 
(provided the operand $L_{1:2,1:2}$ for \ssf{O2} is annotated as weak,) the boundaries between the coarse-grain tasks 
\ssf{O1} and 
\ssf{O2} can be crossed and the execution of \ssf{O2.1} and \ssf{O2.2}
can commence as soon as \ssf{O1.1} is computed.
In order to attain this effect, in OmpSs-2 we should annotate \ssf{O2} as a task with weak operands (via the corresponding representants):
\vspace*{1ex}
\begin{lstlisting}[language=C,caption=,morekeywords={pragma,omp,task}]
#pragma oss task weakin( RepL[0;S] ), weakinout( RepB[0;S] ) 
void task_ltrsm( int M, int P, double *L, int LDL, 
                               double *B, int LDB )
\end{lstlisting}
while \ssf{O2.1}, \ssf{O2.2} are both specified as tasks with strong operands:
\vspace*{1ex}
\begin{lstlisting}[language=C,caption=,morekeywords={pragma,omp,task}]
#pragma oss task in( L[0;M*M] ), inout( B[0;M*P] ) 
void task_ltrsm( int M, int P, double *L, int LDL, 
                               double *B, int LDB )
\end{lstlisting}

This simple example illustrates that the use of weak dependencies and early release can unleash a higher degree of task-parallelism during
the execution of the \H-LU factorization
as, for example, the execution of 
\ssf{O2.1}, \ssf{O2.2} can proceed in parallel with that of 
\ssf{O1.2}--\ssf{O1.5}; and
\ssf{O2.3}, \ssf{O2.4} in parallel with 
\ssf{O1.2}, \ssf{O1.4}, \ssf{O1.5}; 
see Figure~\ref{fig:dependencies_2}.

\section{Numerical Experiments}
\label{sec:experiments}

In this section we first describe the problem setup and target architecture employed in our experiment. Next we 
analyze the concurrency of the 
parallel implementations of the code for the \H-LU factorization in H2Lib.

\subsection{Mathematical problems}
\label{subsec:appl}

The usage of $\mathcal{H}$-matrices often appears in the context of \emph{Boundary Element Methods (BEM)}~\cite{bem}.
The reason for this is that the discretization of boundary integral equations often yields matrices that
are densely populated and have to be stored efficiently, where \hmats come in handy.
There is also a need of constructing efficient preconditioners for this type of equations,
which can be carried out in \H-arithmetic.
In particular, in the experiments in this section we consider integral equations of the form
\begin{equation*}
\int_\Gamma g(x,y)\, u(y)\, d y = f(x), \qquad \text{for almost all } x \in \Omega,
\end{equation*}
where $\Omega$ can be some $d$-dimensional bounded domain for 
$d \in \{1, 2 ,3\}$.
By choosing suitable test-and-trial spaces $\mathcal{U}_h$ and 
$\mathcal{V}_h$, equipped with some bases
$\left( \varphi_i \right), i \in \mathcal{I}$,
and $\left( \psi_j \right), j \in \mathcal{J}$,
we can apply a Galerkin discretization and obtain a variational
formulation of the kind
\begin{equation*}
\int_\Omega v_h(x) \int_\Gamma g(x,y)\, u_h(y)\, dy\, dx = 
   \int_\Omega v_h(x)\, f(x)\, dx\, , 
   \qquad \text{for all } v_h \in \mathcal{V}_h.
\end{equation*}
Employing finite element basis functions for these spaces we directly 
obtain a system of linear equations
\begin{equation*}
G u = f,
\end{equation*}
where all the entries of the matrix
\begin{equation*}
g_{ij} = \int_\Omega \varphi_i(x) \int_\Gamma g(x,y)\, \psi_j(y)\, dy\, dx\, ,
\qquad \text{for all } i \in \mathcal{I}, j \in \mathcal{J},
\end{equation*}
are non-zero.

In particular, we consider the Laplace equation in 
$d \in \{1,2,3\}$ dimensions. In these cases the underlying kernel functions
are
\begin{equation*}
g: \mathbb{R}^d \times \mathbb{R}^d \to \mathbb{R}\,, \quad 
g(x,y) = \begin{cases}
         -\log| x - y| &: d = 1,\\
         -\frac{1}{2 \pi}\log\| x - y\|_2 &: d = 2,\\
         \frac{1}{4 \pi}\| x - y\|_2^{-1} &: d = 3.
         \end{cases}
\end{equation*}

For the construction of low-rank blocks within our experiments, we choose
the analytical method of tensor-interpolation \cite{H2Interpolation},
which is applicable in all dimensions. For the sake of lighter storage 
requirements and faster setup times of the $\mathcal{H}$-LU, we further
re-compress all low-rank blocks by a fast singular value decomposition (SVD)~\cite{GVL3}.

\subsection{Setup}

All the experiments in this section were performed using {\sc ieee} 754 double-precision arithmetic, on a single
node of the MareNostrum~4 system at Barcelona Supercomputing Center. The node contains
two Intel Xeon Platinum 8160 sockets, with 24~cores per socket, and
96~Gbytes per of DDR4 RAM.
In Turbo frequency mode  (3.7~GHz),
the theoretical peak performance for a single core is 59.2~GFLOPS (billions of floating-point
operations per second) when using AVX2 instructions. This rate
is reduced to 33.6~GFLOPS when using a single core running in the base frequency (2.1~GHz).
At this point we note that 
the aggregated (theoretical) peak performance of this machine is a linear function of
the operation frequency which, in turn, depends on the specific type of vector instructions that are executed 
(AVX, AVX2, AVX-512) and the number of active cores~\cite{xeonplatinum}.

In the experiments we employed {\tt gcc} 4.8.5, Intel MKL 2017.4 (with AVX2 instructions enabled), and OmpSs-2 ({\tt mcxx} 2.1.0).

\subsection{Matrix-matrix multiplication}

Our first experiment is designed to assess the performance of the 
implementation of the matrix-matrix multiplication routine ({\tt dgemm})
in Intel MKL. This is relevant because it offers an upper bound of the actual performance that can
be obtained from the \H-LU factorization of a hierarchical matrix. This bound will be tight
in case most of the blocks involved in the decomposition are dense and the fragmentation
of the blocks implicit to the matrix hierarchy is not too fine-grained.

\begin{figure}[ht]
\centering
\includegraphics[scale=0.5]{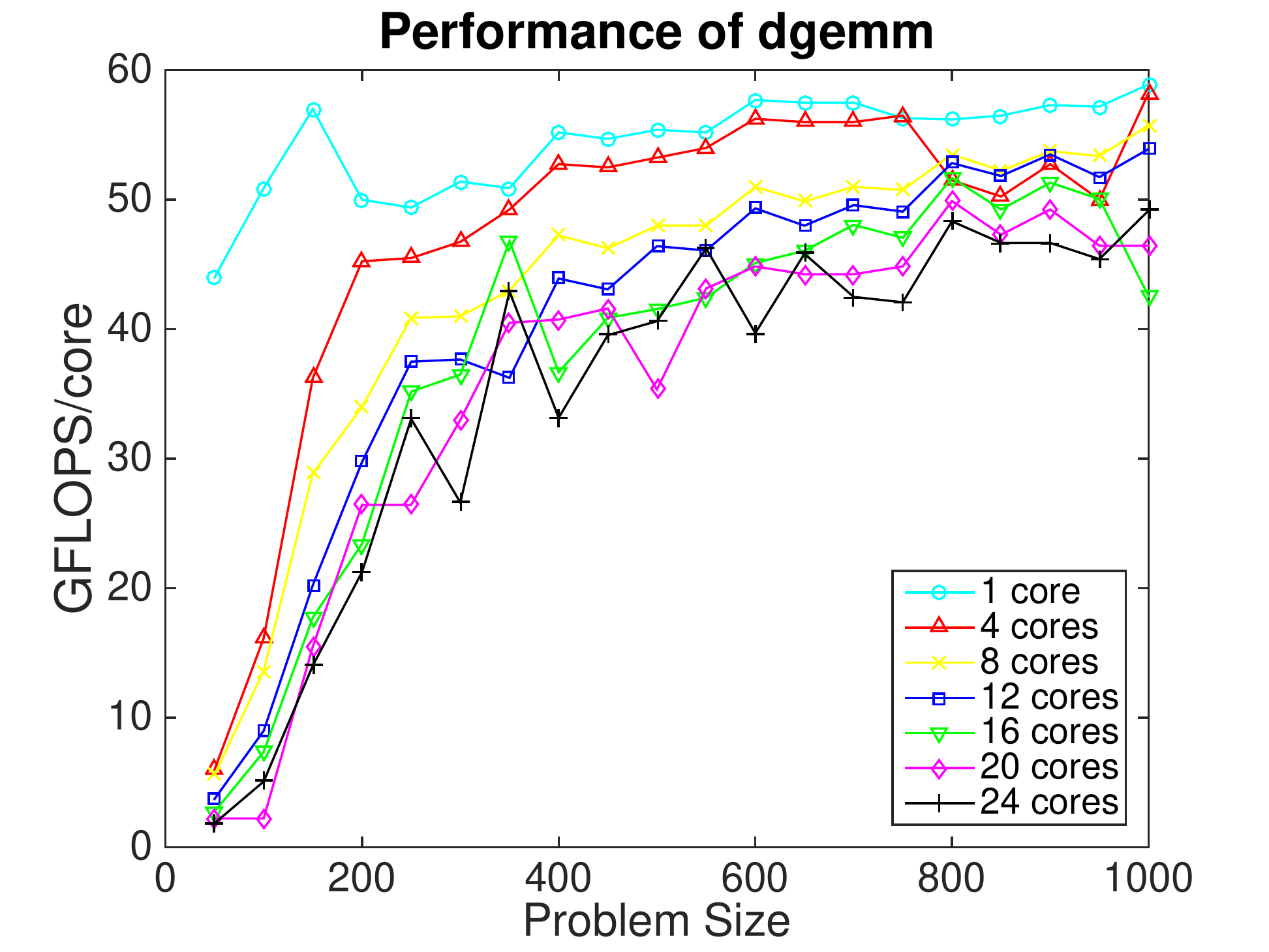}
\caption{Performance of the matrix-matrix multiplication routine in Intel MKL.}
\label{fig:gemm}
\end{figure}

Figure~\ref{fig:gemm} reports the GFLOPS {\em per core} attained by Intel's {\tt dgemm} routine 
using 1, 4, 8,\ldots, 24 cores (of a single socket) and square operands all of the same dimension $b$.
(Note that the limit of the $y$-axis in this plot and all subsequent ones is fixed to 60, which basically corresponds to the
theoretical peak performance with 1 core.)
This experiment reveals two important aspects.
First, the execution of the sequential instance of {\tt dgemm} 
delivers 57.0~GFLOPS for a problem of order
$b=150$, and 58.9~GFLOPS for the largest problem dimension, $b=1000$. 
These values represent 96.2\% an 99.4\% of the peak rate, respectively (when using AVX2 instructions).
Thus, even for problems that are rather small, it is already possible to attain a large fraction of the peak
performance when using a single thread. 
Second, as the number of threads/cores grows, the multi-threaded instance of
{\tt dgemm} requires considerably larger problems to attain a relevant fraction of the theoretical peak.
(As argued earlier, the peak rate of this processor is ``variable'' because it depends on the operation frequency and this
parameter is constrained by the number of active cores~\cite{xeonplatinum}.)

\subsection{Basic parallel solutions}

The following experiment exposes the drawback of a parallelization that simply relies on a multi-threaded instance of the
BLAS, providing initial evidence that a runtime-based approach can offer higher performance.
In order to do so, we compare three different parallelization strategies applied to the \H-LU factorization:
\begin{itemize}
\item {\sf MKL} extracts fine-grain loop-parallelism from within the BLAS kernels only. 
      As argued in the introduction of this paper, this approach is
      rather appealing in that it requires a low programming effort. In particular, provided the sequential routine
      for the \H-LU factorization already casts most of its operations in terms of BLAS, the code can be executed in
      parallel by simply linking in a multi-threaded instance of this library such as that in Intel MKL. 
      The downside of this approach is that it constrains the parallelism that can be leveraged to that
      inside individual kernels, which may be insufficient if the number of cores is large.
\item {\sf OpenMP} aims to exploit loop-parallelism (like {\sf MKL}) but targets a coarser-grain layer, by applying the
      parallelization to the loops present in the \H-LU routine. To clarify this, consider for example a single-level hierarchical
      matrix that is decomposed into $8 \times 8$ blocks. After the factorization of the leading block of the matrix, this
      approach will compute in parallel the remaining 7+7 triangular system solves in the same column+row of the matrix;
      and next update the $7 \times 7$ blocks of the trailing submatrix in parallel. In summary, instead of extracting
      the parallelism from within the individual BLAS kernels, this approach targets the parallelism existing between independent
      BLAS kernels (tasks) comprised by a loop.
\item {\sf OmpSs} discovers tasks dynamically and takes into account the dependencies
      among them to schedule their execution when appropriate; see Section~\ref{sec:pdsec}. 
      (This version does not include the advanced features supported by OmpSs-2.)
\end{itemize}

\begin{figure}[ht]
\centering
\includegraphics[width=0.4\columnwidth]{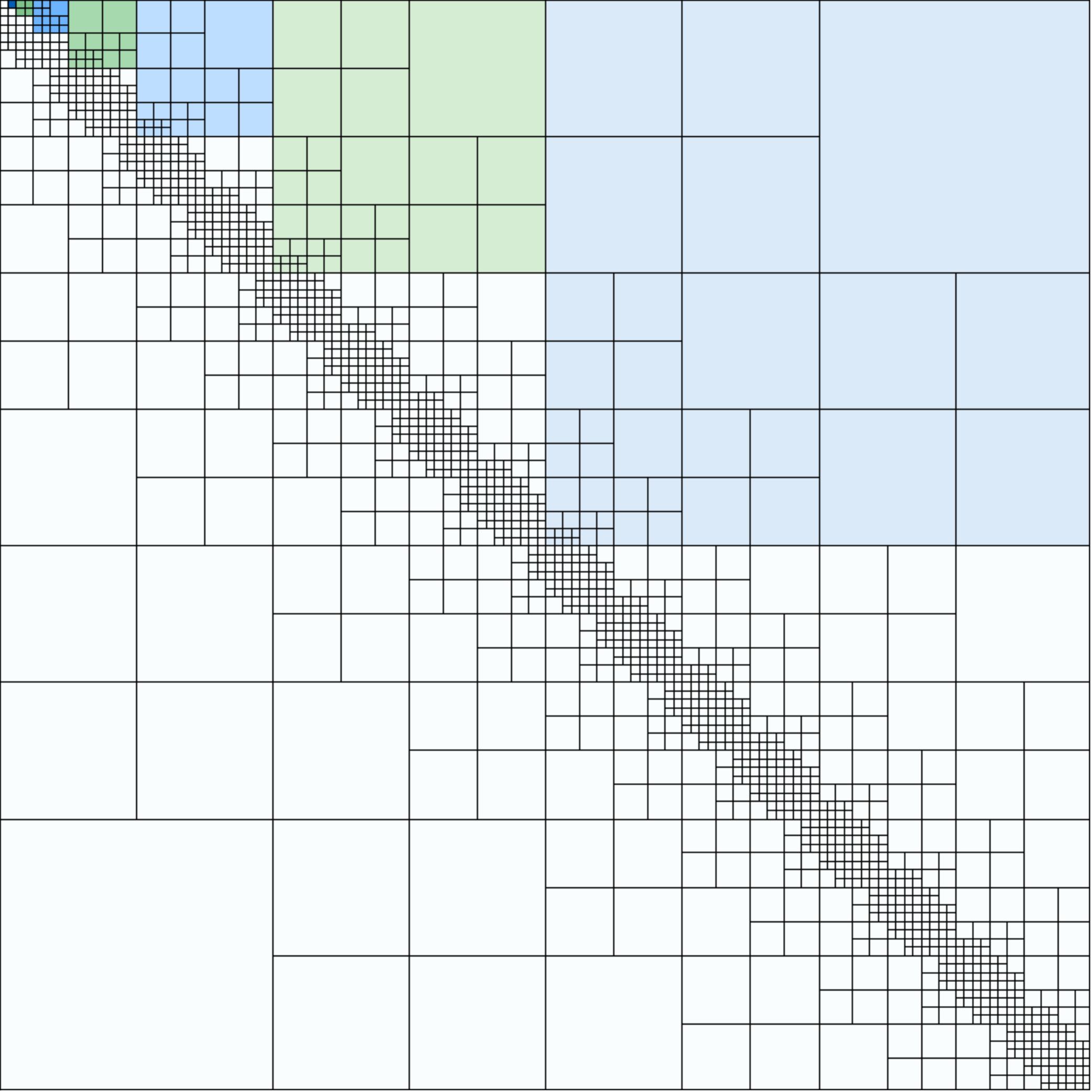} $ $
\includegraphics[width=0.4\columnwidth]{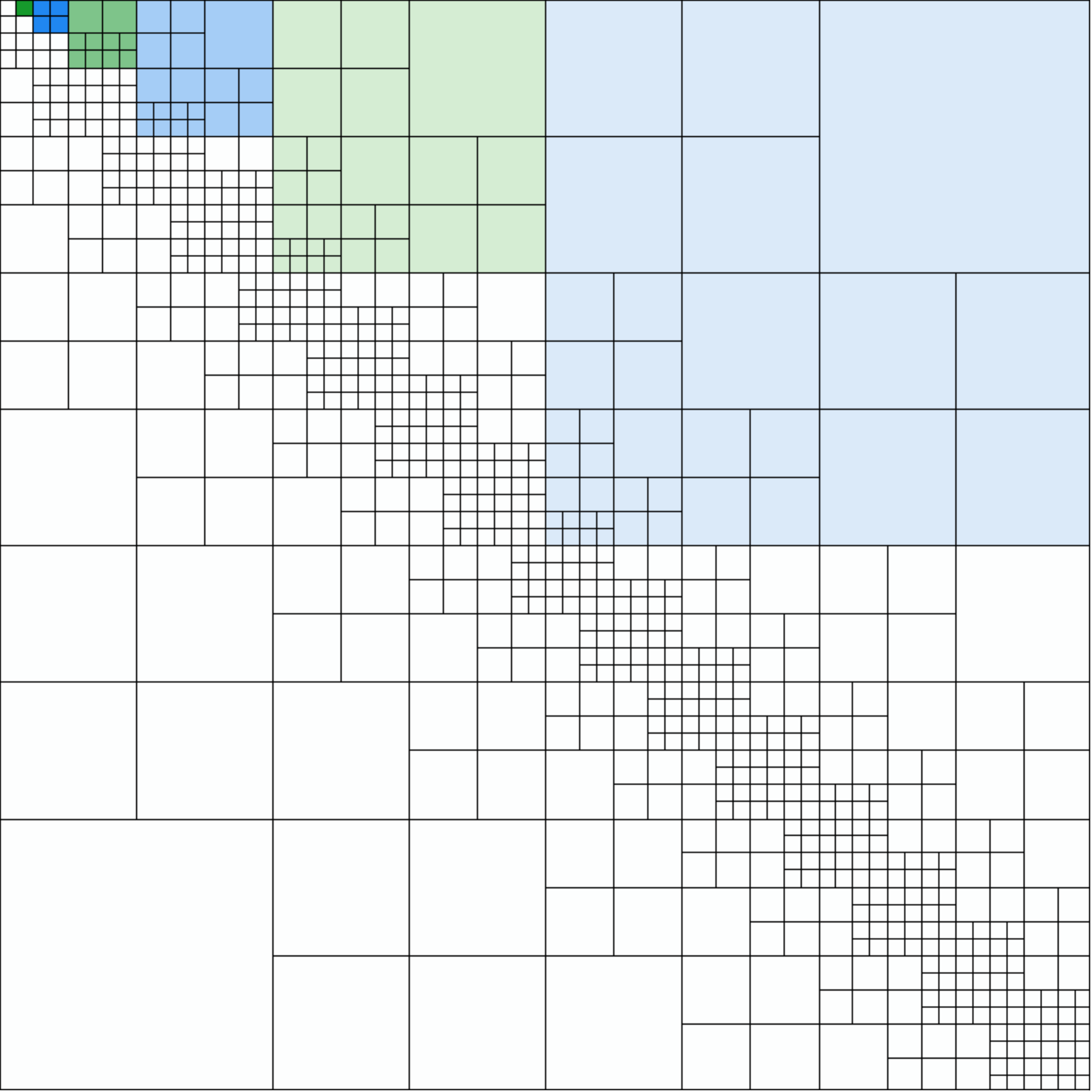}\\ $ $ \\
\includegraphics[width=0.4\columnwidth]{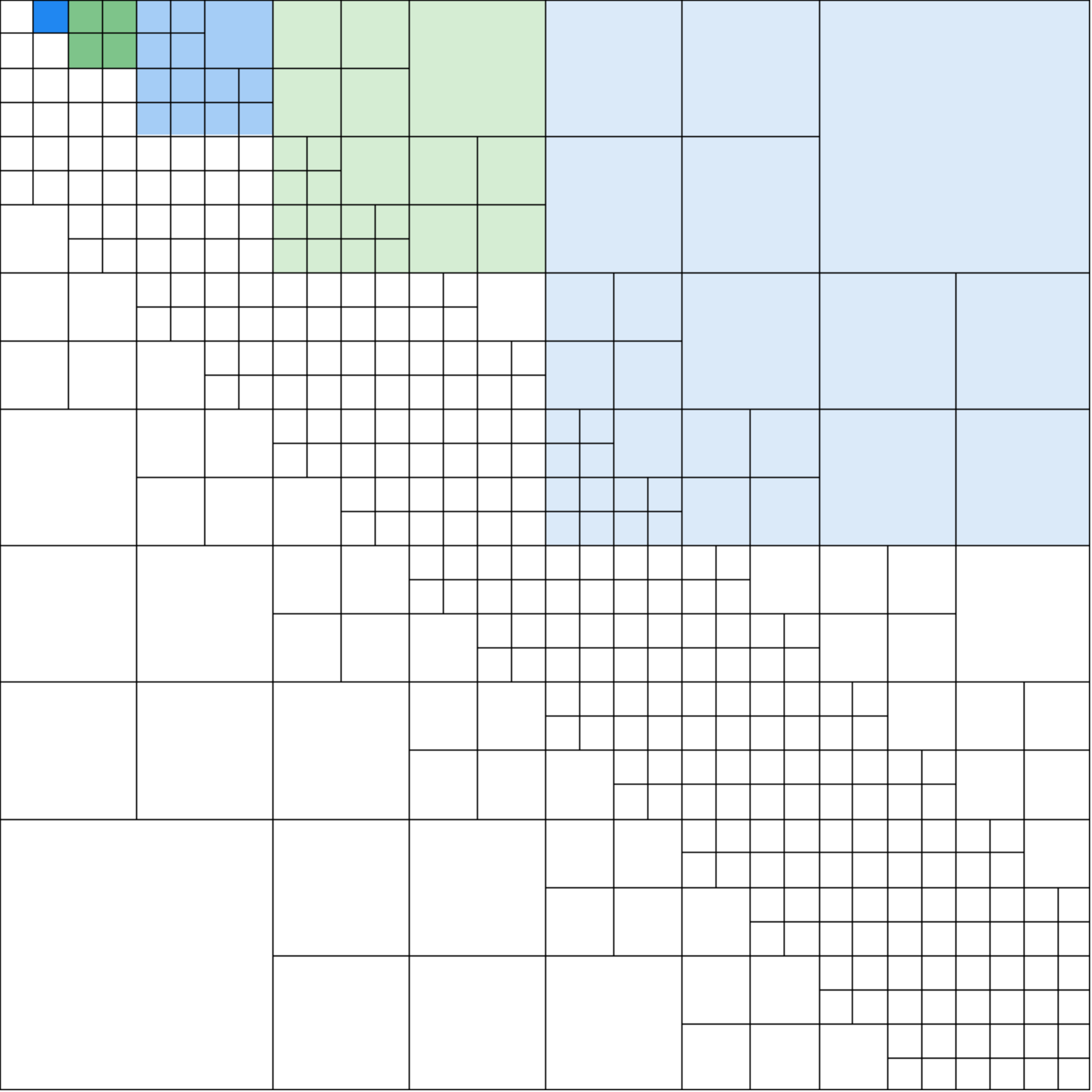} $ $
\includegraphics[width=0.4\columnwidth]{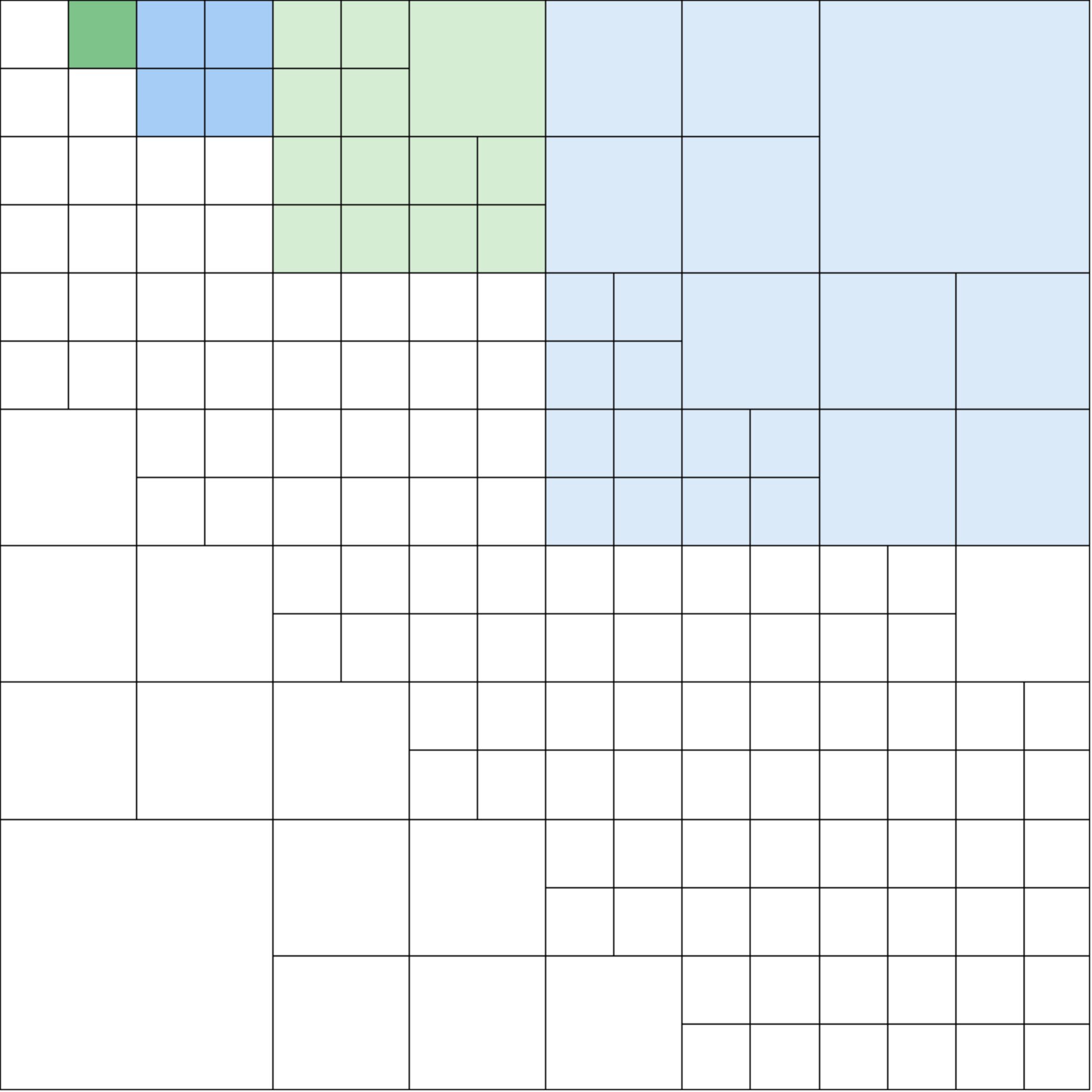}
\caption{Hierarchical structures of the \hmats employed in the evaluation of the parallelization strategies 
        (with 5~recursive partitionings of the diagonal blocks; note that the colors remark the amount of levels defined in each structure: 7, 6, 5 and 4 respectively).}
\label{fig:hmat_blocked_multilevel}
\end{figure}

To simplify the following analysis, we will employ a hierarchical matrix
with a $2\times 2$ recursive structure defined on the diagonal blocks.
Concretely, starting with a hierarchical matrix of order $n$, we define a
$2\times 2$ partitioning, which is recursively applied to the inadmissible blocks until a minimum leafsize is reached;
see Figure~\ref{fig:hmat_blocked_multilevel}.
This type of data structure appears, for example, in BEM with $d=1$ as those described in subsection~\ref{subsec:appl}.
For simplicity, we will also consider dense blocks only.
With these consideration, the cost of the LU factorization of a hierarchical matrix of order $n$ is (approximately) the standard $2n^3/3$ flops.

\begin{figure}[ht]
\centering
\includegraphics[width=0.48\columnwidth,height=5cm]{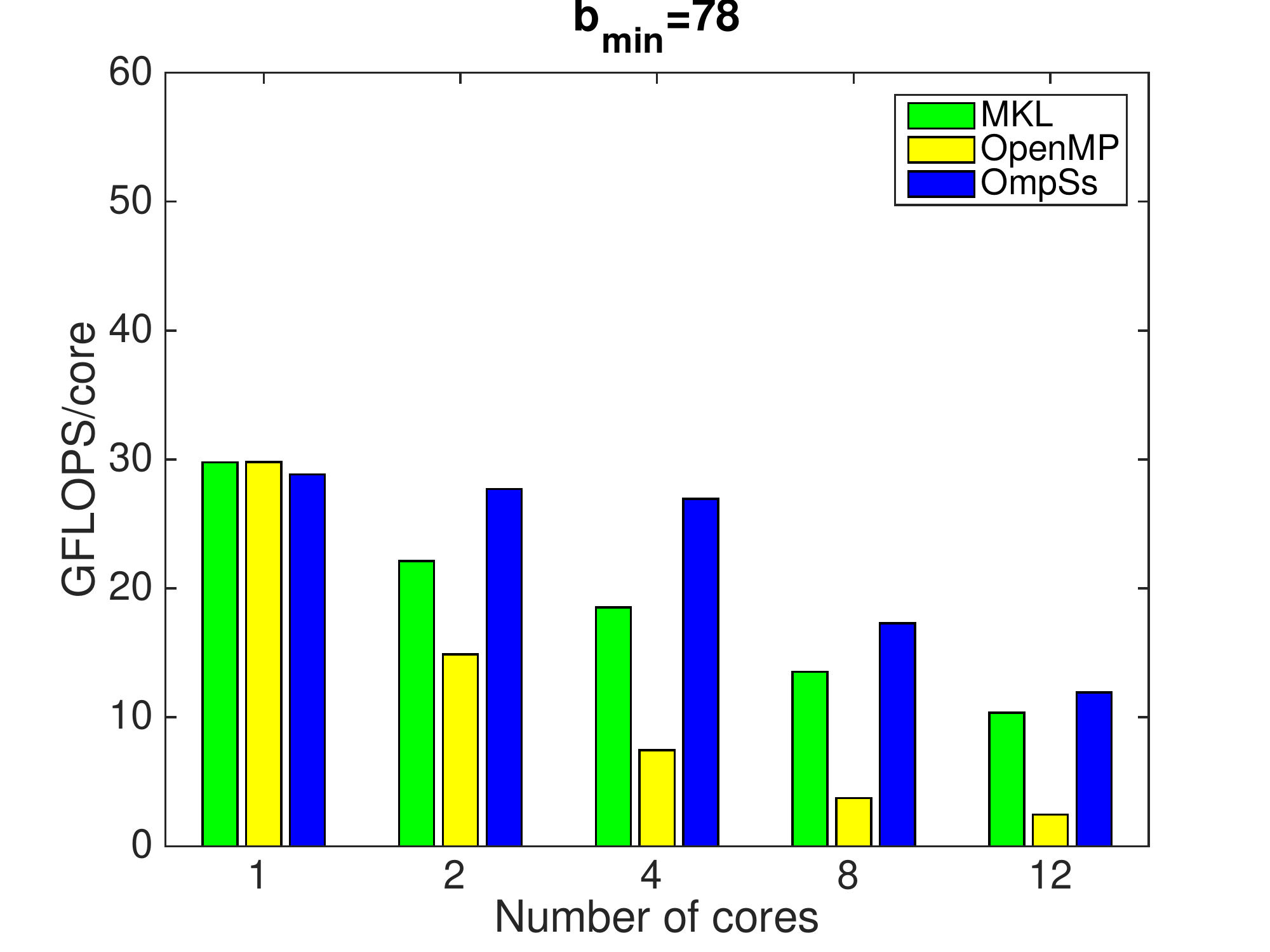}
\includegraphics[width=0.48\columnwidth,height=5cm]{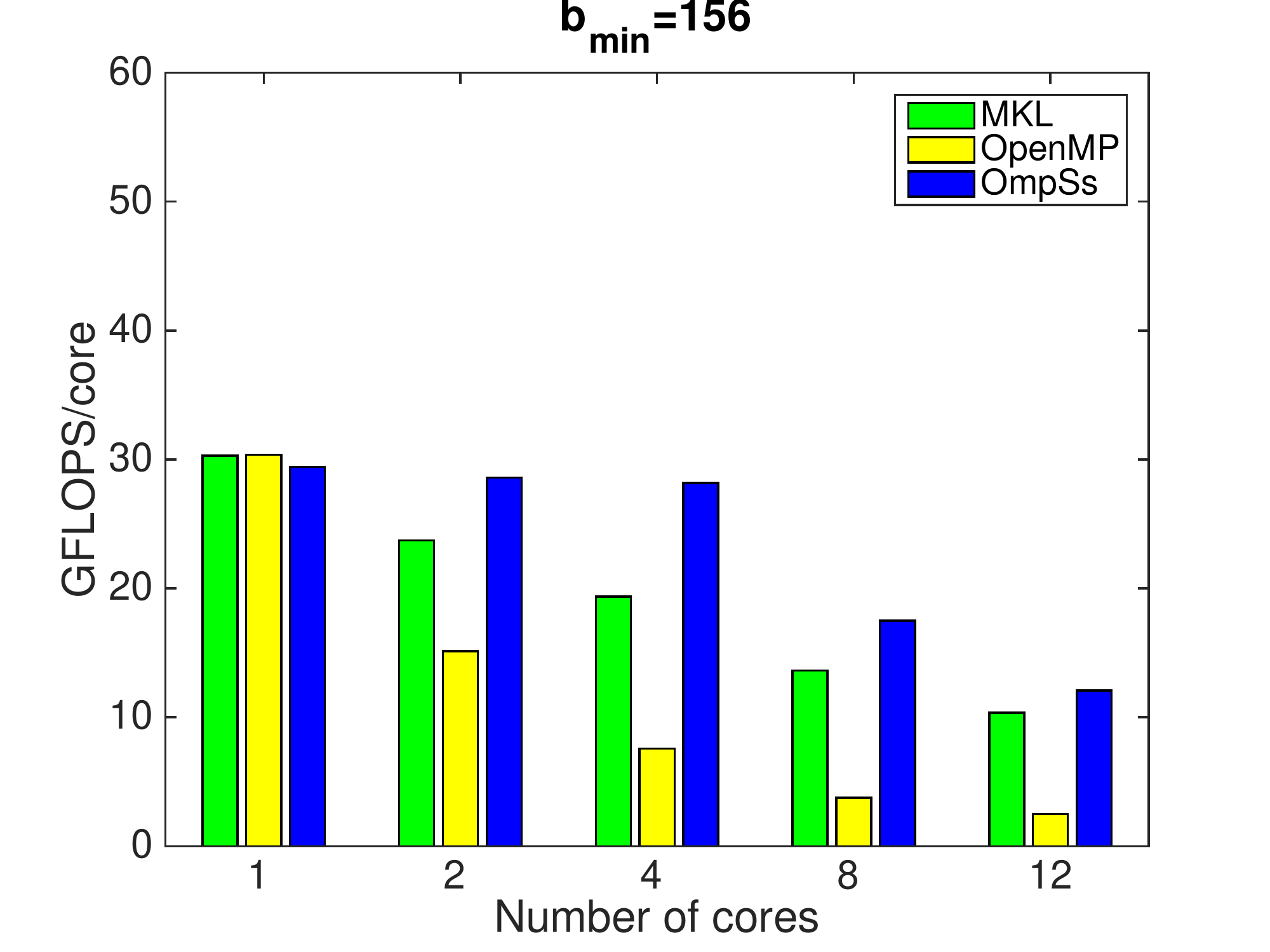}\\
\includegraphics[width=0.48\columnwidth,height=5cm]{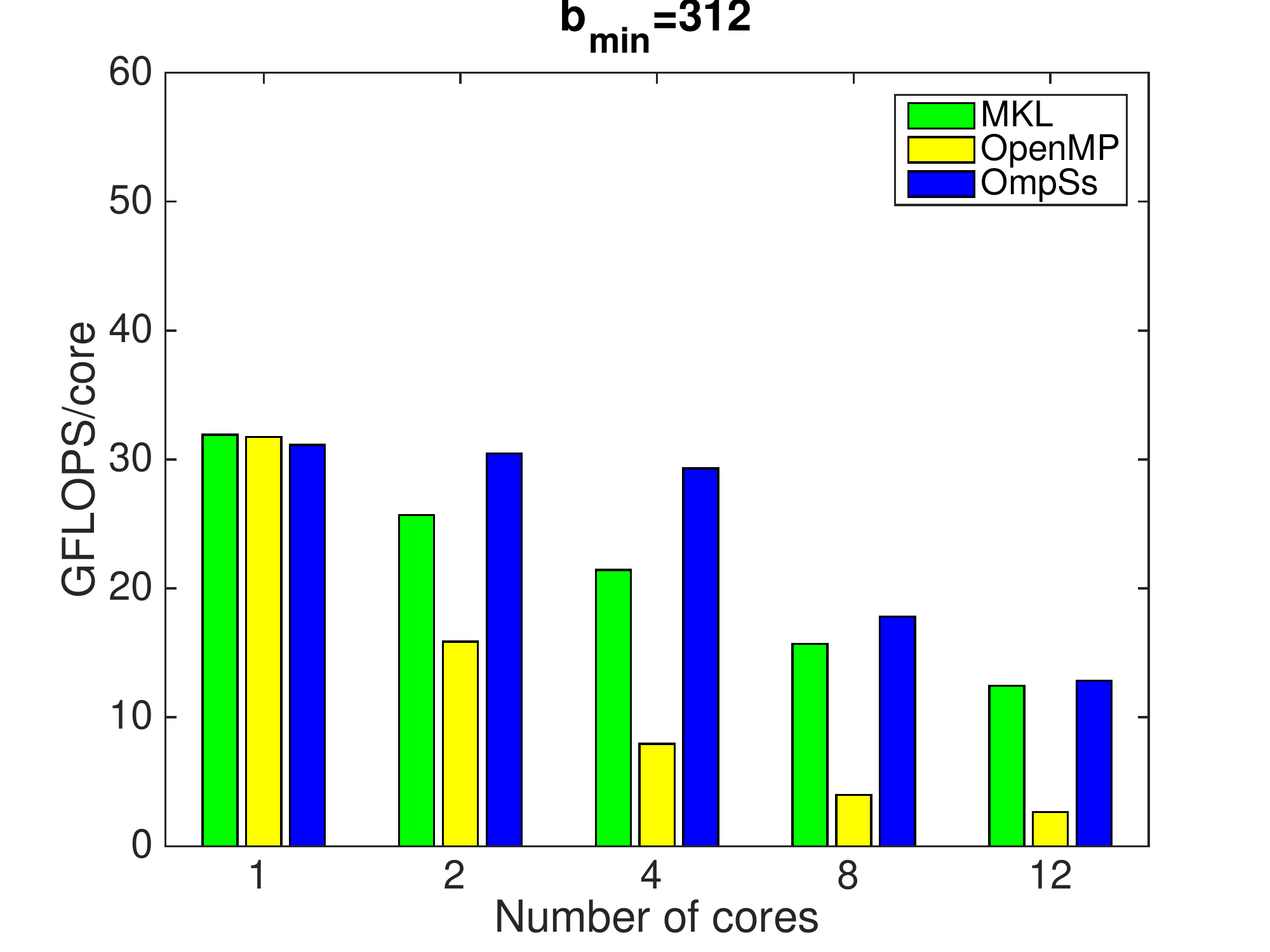}
\includegraphics[width=0.48\columnwidth,height=5cm]{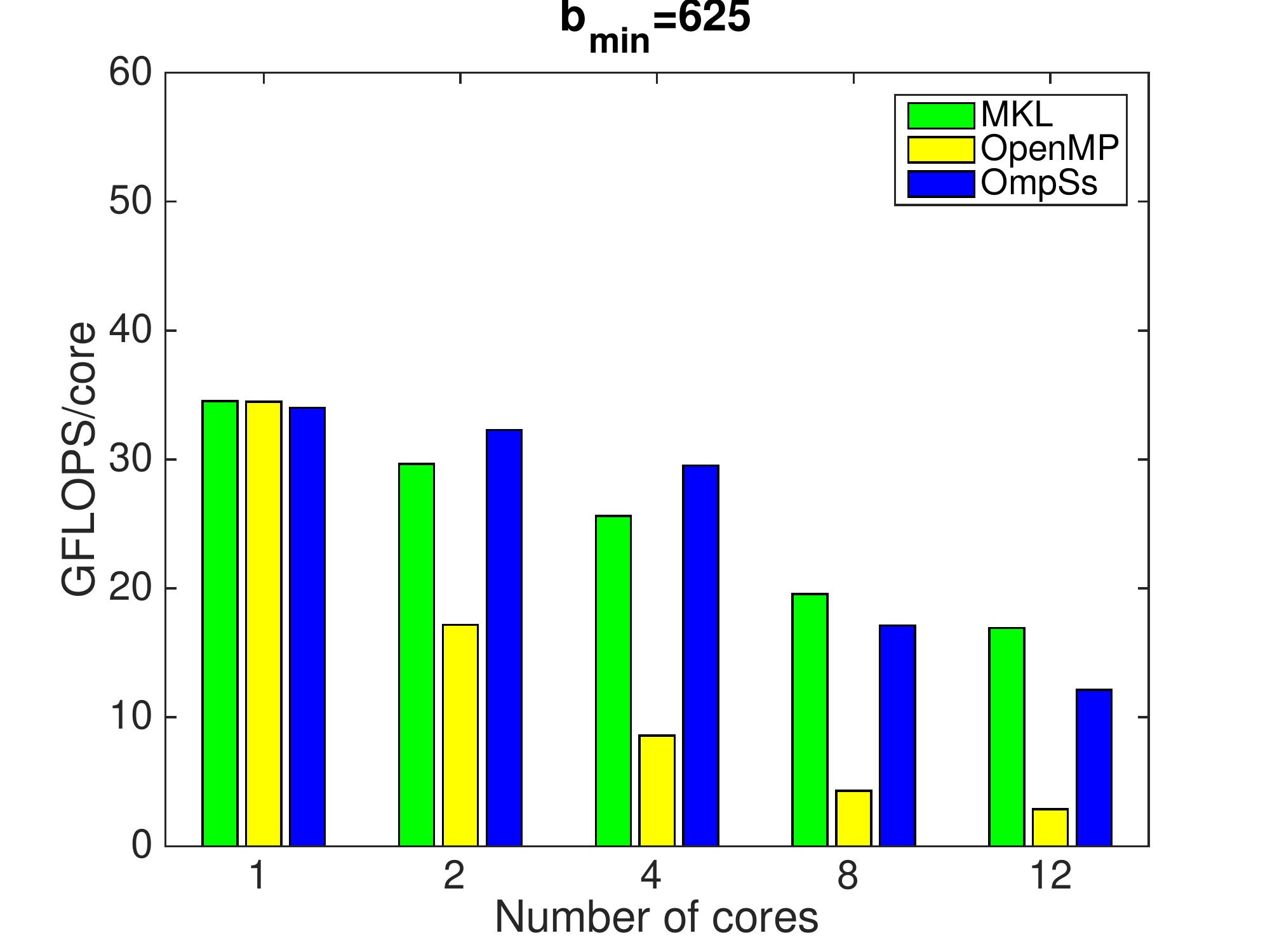}
\caption{Performance of basic parallelization strategies applied to an \hmat of order $n=10$K, with dense blocks, and 
         a recursive $2 \times2$ hierarchical partitioning of the inadmissible blocks; see Figure~\ref{fig:hmat_blocked_multilevel}.}
\label{fig:parallel_strategies}
\end{figure}

Figure~\ref{fig:parallel_strategies}
reports the 
GFLOPS per core
for the three different parallelization strategies described above.
The results there correspond to a square \hmat of dimension $n=10$K,
with $r=4$, 5, 6 and~7 recursive partitionings applied to the inadmissible blocks until a minimum leafsize is reached. This implies that
the smallest blocks on the diagonal are of order $b_{\min}=10\mathrm{K}/2^r \approx 625$, 312, 156 and~78, respectively.
This experiment offers some interesting insights:
\begin{itemize}
\item The performance of {\sf MKL} greatly benefits from problems with large block sizes, which is consistent with
      the trends in the GFLOPS rates observed for the multi-threaded instance of Intel's {\tt dgemm} in the previous experiment.
      This option is competitive
      with the task-parallel OmpSs-based routine when the number of cores is reduced or partitioning features large diagonal blocks ($r=4$, $b_{\min}=625$).
\item The parallel performance of {\sf OpenMP} is practically negligible as the GFLOPS per core decrease linearly with the number of 
      cores. 
      This is not totally a surprise as, due to the
      $2\times 2$ organization of the \hmat, the operations that can be performed independently are reduced to
      the two triangular system solves at each partitioning.
\item When the number of cores is small, the OmpSs-based parallelization attains
      mild GFLOPS rates. Here, the coarse-grain partitioning of the blocks 
      and the existence of synchronization points constrain the degree of parallelism that can be exploited and limit the performance
      of this approach when the number of cores is large.
\end{itemize}

To complete the analysis of this experiment, we remark that a comparative analysis of the GFLOPS observed in these executions
with those of {\tt dgemm} is delicate. In particular, the execution using a single core can be expected to 
set the processor to operate on a higher frequency than
a parallel multi-threaded execution using several cores.
Unfortunately, the exact frequency is difficult to know as it depends on the number of cores as well as the arithmetic intensity of the 
operations (and it can even vary at execution time). 

\subsection{Scalability of task-parallel routines}

Our next experiments aim to demonstrate the benefits that the 
{\sf WD+ER} (weak dependency and early release) mechanism exerts on the scalability of the task-parallel codes based on OmpSs-2.
For this purpose, we next conduct an analysis of the strong and weak scalabilies,
using a complete socket (24 cores) 
and the same hierarchical matrix, with a $2\times 2$ recursive structure defined on the diagonal blocks and dense blocks only, employed in the previous study.

\begin{figure}[ht]
\centering
\includegraphics[scale=0.5]{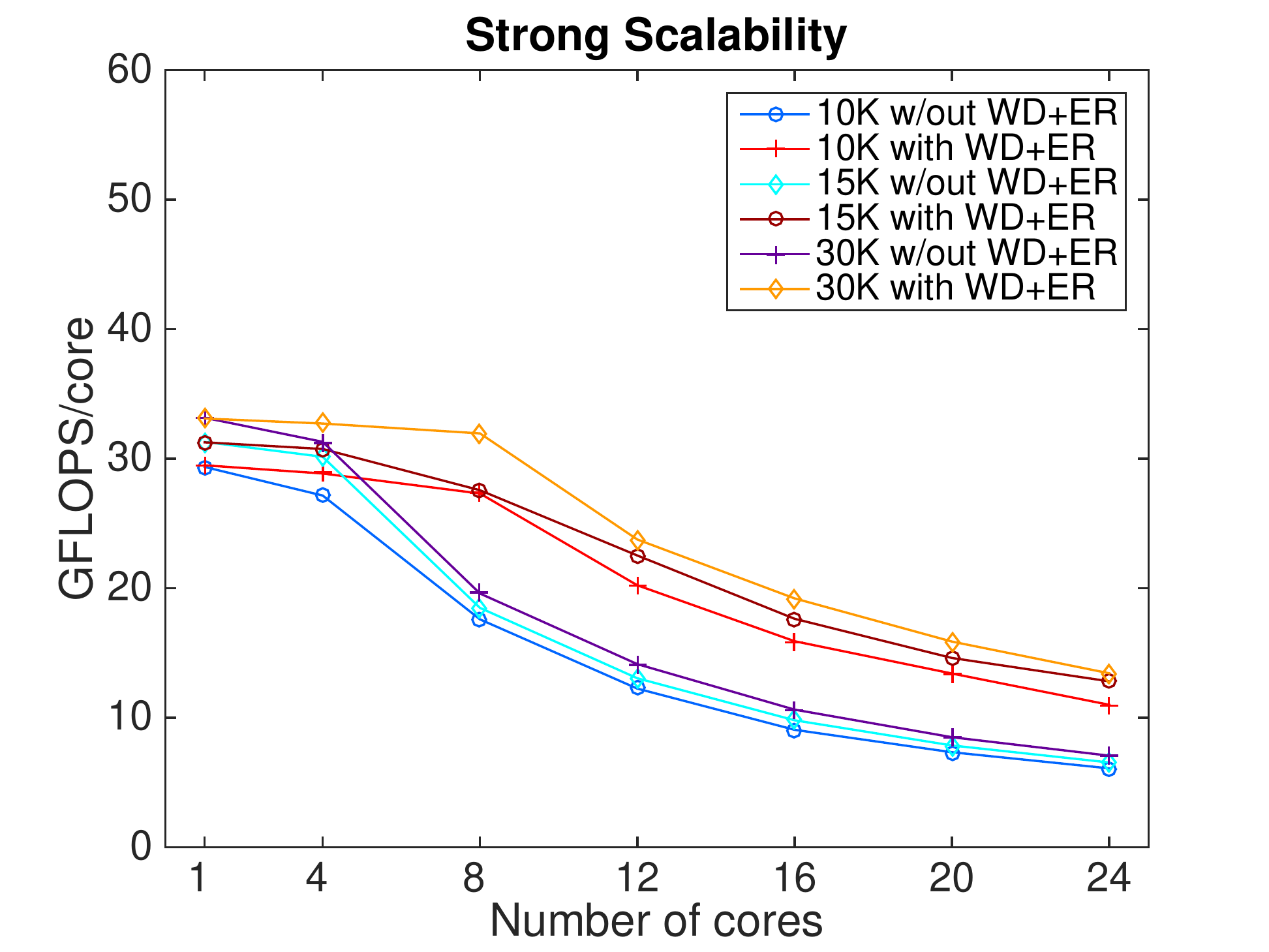}
\caption{Strong scalability of the advanced parallelization strategies applied to \hmats of order $n=10$K, 15K and 30K
         ($b_{\min}=$156, 234 and 234, respectively), with dense blocks, and 
         a recursive $2 \times2$ hierarchical partitioning of the inadmissible blocks; see Figure~\ref{fig:hmat_blocked_multilevel}.}
\label{fig:strong_strategies}
\end{figure}

In the following analysis of strong scalability, we set the problem dimension to three different values, $n=$10K, 15K and 30K, and progressively increase the amount of cores up
to 24 while measuring the GFLOPS per core. In this type of experiment, we can expect that the GFLOPS/core rates eventually drop as the problem becomes too 
small for the volume of resources that are employed to tackle it. 
Figure~\ref{fig:strong_strategies} confirms that this is the case for both implementations, which exploit/do not exploit the new features
in OmpSs-2 (lines labeled as {\sf with WD+ER} and {\sf w/out WD+ER}, respectively).
In addition, the results also show that the exploitation of {\sf WD+ER}, made possible by OmpSs-2, offers a GFLOPS/core rate that
clearly outperforms that of the implementation that is oblivious of these options.

For the analysis of weak scalability, we utilize a problem of dimension $n \times n$ that grows proportionally to the number of cores $c$, so that the
ratio $n^2/c = 15$K$\times15$K holds while $c$ grows to 24.
As the problem size per core is constant, we can expect that the GFLOPS/core
remains stable, showing the possibility of addressing larger problems by increasing proportionally the amount of resources up to a certain point.
(This is not totally exact, as the cost of the factorization for dense matrices grows cubically with the problem dimension while, in the conditions
set for this experiment, the amount of resources only does so quadratically.)
Unfortunately, the results of this experiment reveal that the weak scalability of both algorithms suffers an important drop as the number of cores
is increased, though in the variant equipped with {\sf WD+ER} this occurs in the transition from~8 to 12~cores while the implementation that does
not exploit this mechanism the gap is visible already in the increase from~4 to~8 cores.

There are two aspects to take into account when considering the GFLOPS/core rates observed in the strong scaling analysis and, especially, the weak scaling 
counterpart. The first one refers to the CPU frequency, which decreases with the number of cores which are active (see~\cite{xeonplatinum} and
Figure~\ref{fig:gemm}) and affect the performance of the task-parallel routines, reducing it with the number of cores.
The second one is a consideration of the structure of the hierarchical matrix employed in these experiments (see Figure~\ref{fig:hmat_blocked_multilevel}).
In particular, when all the blocks are dense, and the matrix is decomposed into a task per block in this partitioning, the result is a problem where a reduced collection of
coarse-grain tasks concentrate a large fraction of the flops. This effect is exacerbated with the problem order ($n$) and its negative effect is more visible
when the number of cores is increased because the task-parallel algorithms confront then a suboptimal scenario consisting of a very reduced number of tasks
(little task-parallelism) of (very) coarse-grain operations.

\begin{figure}[ht]
\centering
\includegraphics[scale=0.5]{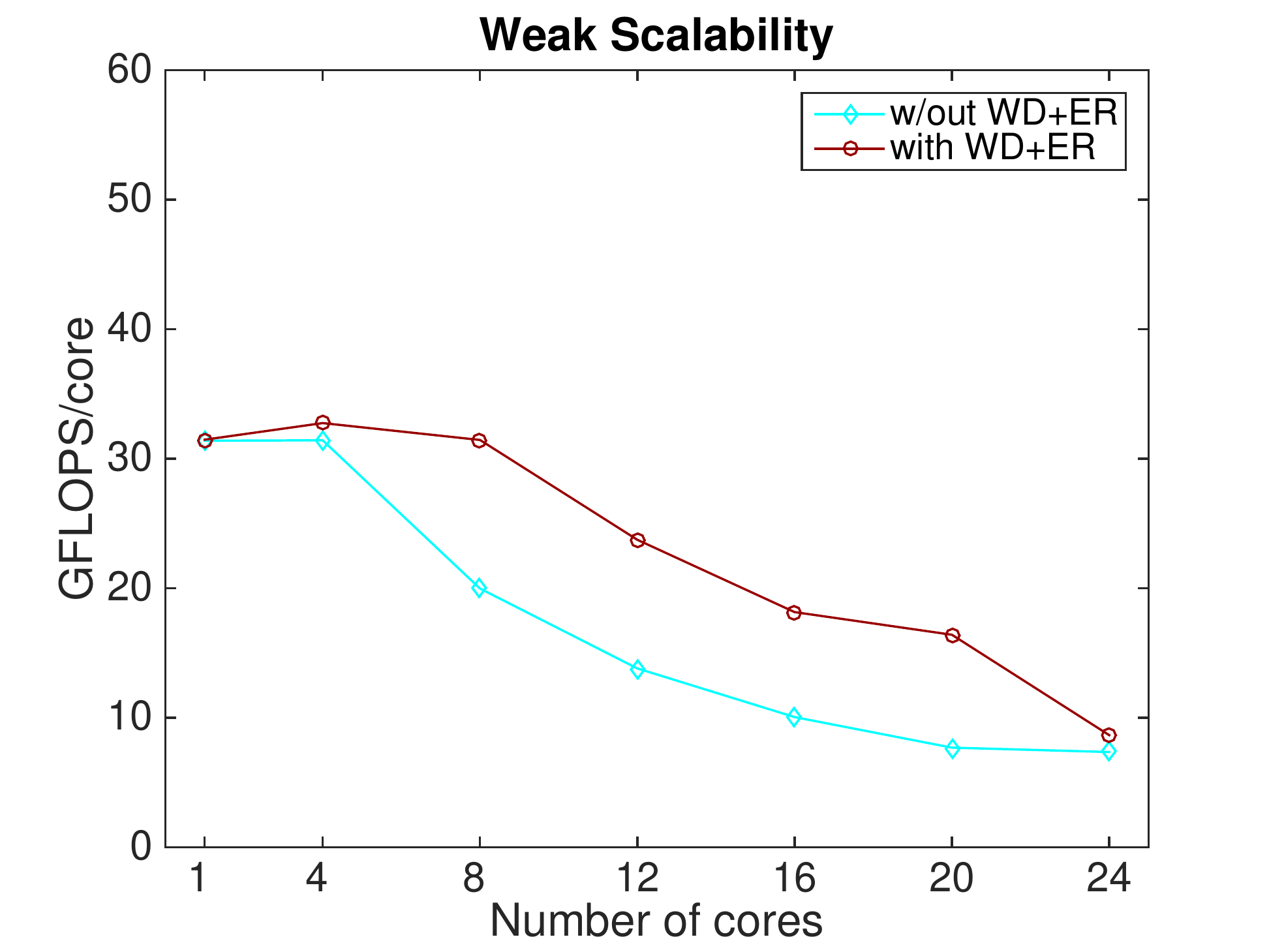}
\caption{Weak scalability of the advanced parallelization strategies applied to an \hmat of dimension $n\times n=15$K$\times15$K per core
         ($b_{\min}=$234 in all cases, except with 8~cores where $b_{\min}=$166), with dense blocks, and 
         a recursive $2 \times2$ hierarchical partitioning of the inadmissible blocks; see Figure~\ref{fig:hmat_blocked_multilevel}.}
\label{fig:weak_strategies}
\end{figure}

\subsection{Parallelism of task-parallel routines with low-rank cases}

\begin{figure}[ht]
\centering
\begin{tabular}{ccc}
\begin{minipage}[t]{0.25\textwidth} \includegraphics[width=\textwidth]{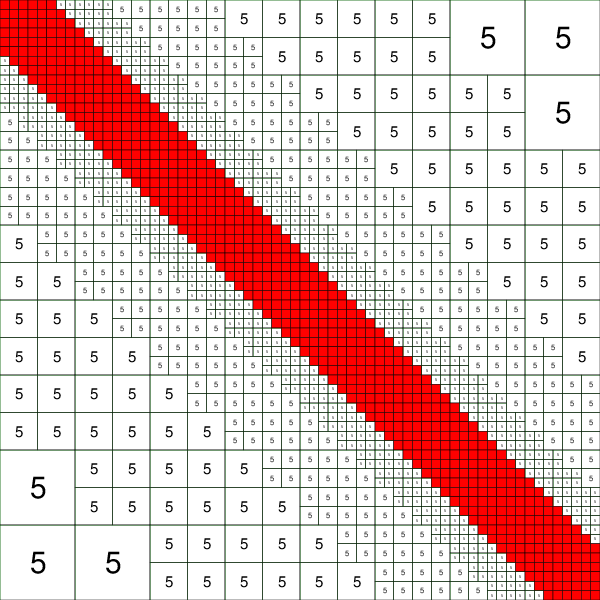} \end{minipage} &
\begin{minipage}[t]{0.25\textwidth} \includegraphics[width=\textwidth]{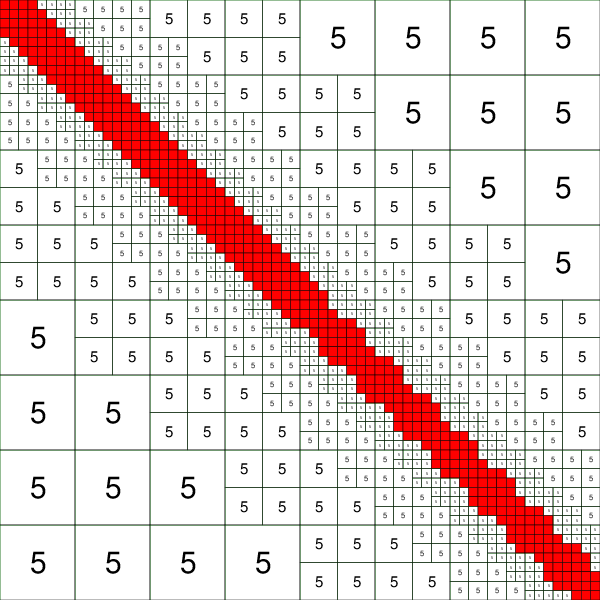} \end{minipage} &
\begin{minipage}[t]{0.25\textwidth} \includegraphics[width=\textwidth]{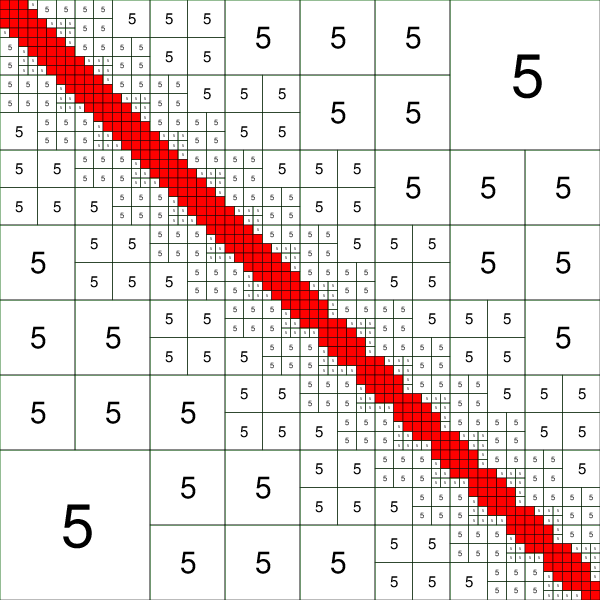} \end{minipage} \\
\begin{minipage}[t]{0.25\textwidth} \includegraphics[width=\textwidth]{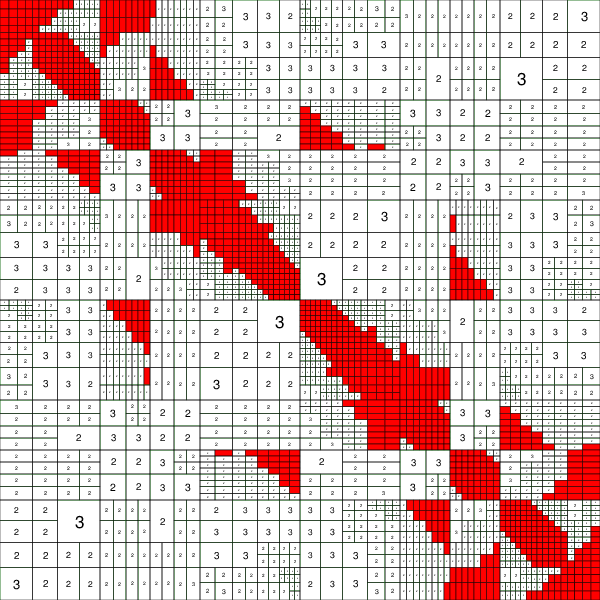} \end{minipage} &
\begin{minipage}[t]{0.25\textwidth} \includegraphics[width=\textwidth]{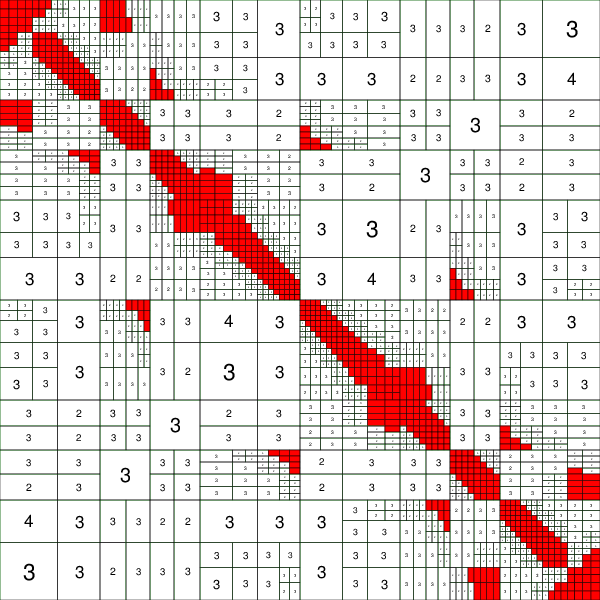} \end{minipage} &
\begin{minipage}[t]{0.25\textwidth} \includegraphics[width=\textwidth]{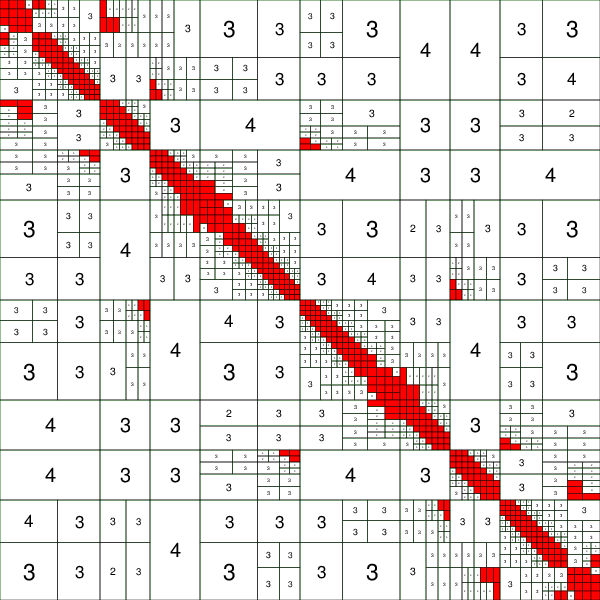} \end{minipage} \\
\begin{minipage}[t]{0.25\textwidth} \includegraphics[width=\textwidth]{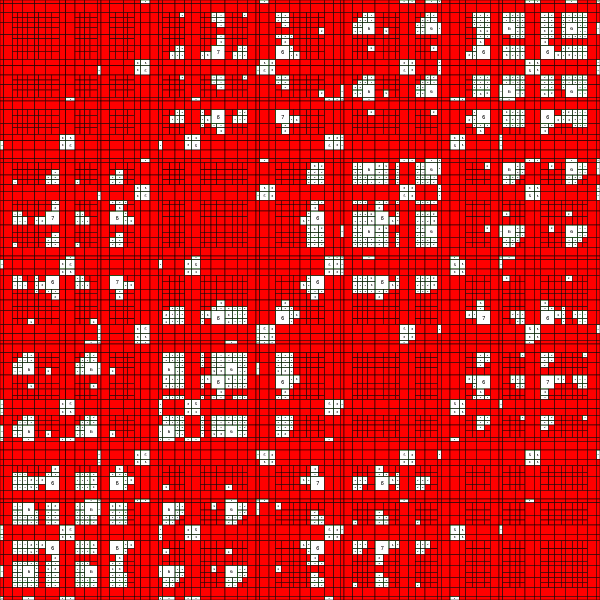} \end{minipage} &
\begin{minipage}[t]{0.25\textwidth} \includegraphics[width=\textwidth]{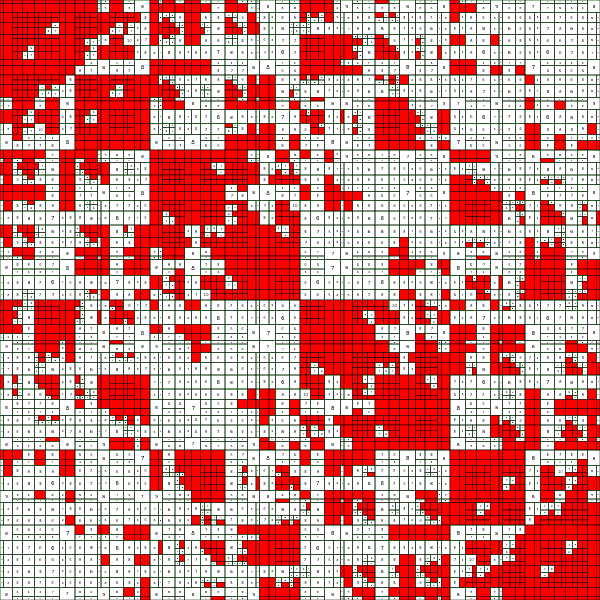} \end{minipage} &
\begin{minipage}[t]{0.25\textwidth} \includegraphics[width=\textwidth]{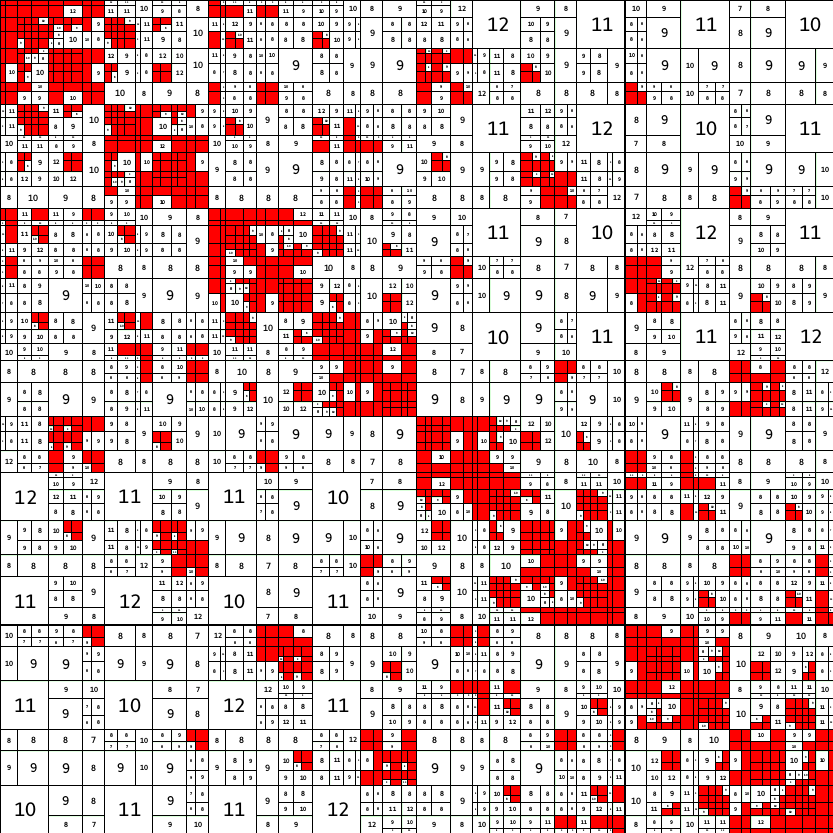} \end{minipage} \\
\end{tabular}
\caption{Hierarchical structure of the \hmat employed in the evaluation of the 
         task-parallel routines. The red areas denote dense blocks and the number inside the white blocks specifies the rank of the 
         corresponding (low-rank) block.
         From top to bottom: $d=$1, 2 and 3; and from left to right: 
         $\eta=$0.25, 0.5, 1.0 
         in the first two rows, and
         $\eta=$0.5, 1.0, 2.0 in the last one.}
\label{fig:hmat_blocked_123d}
\end{figure}

Our final round of experiments assesses the performance of the {\sf WD+ER} (weak dependency and early release) mechanism
using several BEM cases, of dimensions $d=$1, 2 and 3, involving low-rank blocks.
The ``sparsity'' pattern of these blocks is controlled via a parameter $\eta$ that we set to four different values,
0.25, 0.5, 1.0 and 2.0. The structure of these cases is illustrated in Figure~\ref{fig:hmat_blocked_123d}.

Table~\ref{tab:hmat_blocked_123d} reports the acceleration factors (or speed-ups) attained by the task-parallel codes with respect to the corresponding sequential code/case,
using problems of order $n \approx 30$K and up to 24 cores.
This final experiment illustrates the performance advantage of exploiting the {\sf WD+ER} also in case of \hmats with low-rank blocks. 
In general, 
the speed-up increases with the ratio of dense blocks, reporting notably high values for $d=1$ and 3 (provided the number of
cores is not too large compared with the problem dimension), and much lower for
those cases with $d=2$. In some cases we even observe a super-linear speed-up, due to cache effects.

\begin{table}[h!]
\centering
\begin{center}
{\footnotesize
\begin{tabular}{ |l|l|l||c||c|c|c|c|c|c| } 
\hline
         &     & {\sf WD+}                    & Seq.   & \multicolumn{6}{c|}{Speed-up with \#cores} \\
  $\eta$ & $d$ & {\sf ER}?                    & time   & 4       & 8       & 12       & 16       & 20       & 24       \\
\hline \hline
\multirow{4}{*}{0.25} & \multirow{2}{*}{1} & No  & \multirow{2}{*}{~~89.2}       & ~3.51 & ~5.24 & ~5.58 & ~5.70 & ~5.70 & ~5.69 \\
                      &                    & Yes &        & ~4.05 & ~7.82 & 11.56 & 14.62 & 17.37 & 19.04 \\ \cline{2-10}
                      & \multirow{2}{*}{2} & No  & \multirow{2}{*}{~118.9}       & ~3.70 & ~6.12 & ~7.28 & ~7.79 & ~7.88 & ~8.02 \\
                      &                    & Yes &        & ~3.96 & ~7.64 & 10.99 & 13.55 & 17.36 & 18.49 \\ \cline{2-10}
\hline \hline
\multirow{6}{*}{0.50} & \multirow{2}{*}{1} & No  & \multirow{2}{*}{~~38.1}       & ~2.60 & ~2.74 & ~2.73 & ~2.71 & ~2.71 & ~2.70 \\
                      &                    & Yes &        & ~3.92 & ~7.45 & ~9.57 & ~9.74 & ~9.69 & ~9.45 \\ \cline{2-10}
                      & \multirow{2}{*}{2} & No  & \multirow{2}{*}{~~37.0}       & ~3.04 & ~3.68 & ~3.82 & ~3.90 & ~5.74 & ~3.92 \\
                      &                    & Yes &        & ~3.96 & ~6.87 & ~8.81 & ~9.68 & 10.18 & 10.44 \\ \cline{2-10}
                      & \multirow{2}{*}{3} & No  & \multirow{2}{*}{1,099.2}       & ~4.00 & ~7.83 & 13.48 & 14.44 & 16.88 & 18.71 \\
                      &                    & Yes &        & ~4.00 & ~7.99 & 11.85 & 15.63 & 19.13 & 21.57 \\ \hline \hline
\multirow{6}{*}{1.00} & \multirow{2}{*}{1} & No  & \multirow{2}{*}{~~12.6}       & ~1.55 & ~1.58 & ~1.57 & ~1.57 & ~1.56 & ~1.57 \\
                      &                    & Yes &        & ~2.50 & ~2.46 & ~2.48 & ~2.46 & ~2.44 & ~2.43 \\ \cline{2-10}
                      & \multirow{2}{*}{2} & No  & \multirow{2}{*}{~~12.0}       & ~1.92 & ~1.99 & ~2.00 & ~2.00 & ~1.99 & ~2.00 \\
                      &                    & Yes &        & ~3.31 & ~4.22 & ~4.38 & ~4.50 & ~4.44 & ~4.36 \\ \cline{2-10}
                      & \multirow{2}{*}{3} & No  & \multirow{2}{*}{1,049.8}       & ~3.96 & ~7.54 & 10.84 & 13.28 & 15.64 & 17.63 \\
                      &                    & Yes &        & ~4.03 & ~7.91 & 14.32 & 15.55 & 18.99 & 21.73 \\
\hline \hline
\multirow{2}{*}{2.00} & \multirow{2}{*}{3} & No  & \multirow{2}{*}{~204.1}       & ~3.59 & ~5.78 & ~7.39 & ~7.88 & ~8.26 & ~8.44 \\
                      &                    & Yes &        & ~4.99 & ~7.87 & 11.27 & 14.65 & 17.33 & 17.83 \\ \cline{2-10}
\hline
\end{tabular}
}
\end{center}
\caption{Execution time of the sequential algorithm in H2Lib (in sec.) and parallel 
         speed-up of the advanced parallelization strategies applied to an \hmat of order $n\approx30$K
         ($b_{\min}=$234), with dense and low-rank blocks; see Figure~\ref{fig:hmat_blocked_123d}.}
\label{tab:hmat_blocked_123d}
\end{table}


\section{Concluding Remarks}
\label{sec:remarks}

We have demonstrated notable parallel efficiency for the calculation of the \H-LU factorization on a 
state-of-the-art Intel Xeon socket with 24 cores. A key component to attain this high performance is
the exploitation of weak dependencies and early release, recently introduced in OmpSs-2. Armed with these mechanisms, 
the OmpSs-based parallel codes can cross the dependency domains, discovering and exploiting a notably
higher degree of task-parallelism, which results in higher performance in the execution 
of 1D, 2D and 3D cases arising from BEM. As part of future work, we would like to investigate 
hybrid parallelization schemes that combine the extraction of multi-threaded parallelism from 
highly tuned libraries such as Intel MKL with task-parallelism exploited by a runtime. Moreover, we will investigate new strategies to extract additional levels of task-parallelism.


\section*{Acknowledgments}

The researchers from Universidad Jaume~I (UJI)
were supported by projects CICYT TIN2014-53495-R and TIN2017-82972-R of
MINECO and FEDER; project UJI-B2017-46 of UJI; and the FPU program of MECD.

\bibliographystyle{elsarticle-num} 
\bibliography{biblio}

\begin{thebibliography}{10}
\expandafter\ifx\csname url\endcsname\relax
  \def\url#1{\texttt{#1}}\fi
\expandafter\ifx\csname urlprefix\endcsname\relax\def\urlprefix{URL }\fi
\expandafter\ifx\csname href\endcsname\relax
  \def\href#1#2{#2} \def\path#1{#1}\fi

\bibitem{Hackbusch:1999:SMA:303815.303816}
W.~Hackbusch, \href{http://dx.doi.org/10.1007/s006070050015}{A sparse matrix
  arithmetic based on $\mathcal{H}$-matrices. {P}art {I}: Introduction to
  $\mathcal{H}$-matrices}, Computing 62~(2) (1999) 89--108.
\newblock \href {http://dx.doi.org/10.1007/s006070050015}
  {\path{doi:10.1007/s006070050015}}.
\newline\urlprefix\url{http://dx.doi.org/10.1007/s006070050015}

\bibitem{Hack09}
W.~Hackbusch, Hierarchical Matrices: Algorithms and Analysis, Vol.~49 of
  Springer Series in Computational Mathematics, Springer-Verlag Berlin
  Heidelberg, 2015.

\bibitem{GrasHack:Arithm}
L.~Grasedyck, W.~Hackbusch,
  \href{http://dx.doi.org/10.1007/s00607-003-0019-1}{Construction and
  arithmetics of $\mathcal{H}$-matrices}, Computing 70~(4) (2003) 295--334.
\newblock \href {http://dx.doi.org/10.1007/s00607-003-0019-1}
  {\path{doi:10.1007/s00607-003-0019-1}}.
\newline\urlprefix\url{http://dx.doi.org/10.1007/s00607-003-0019-1}

\bibitem{BLAS3}
J.~J. Dongarra, J.~Du~Croz, S.~Hammarling, I.~Duff, A set of level 3 basic
  linear algebra subprograms, ACM Trans. on Mathematical Software 16~(1) (1990)
  1--17.

\bibitem{Buttari200938}
A.~Buttari, J.~Langou, J.~Kurzak, , J.~Dongarra, A class of parallel tiled
  linear algebra algorithms for multicore architectures, Parallel Computing
  35~(1) (2009) 38--53.

\bibitem{Quintana:2008:PMA}
G.~Quintana-Ort{\'\i}, E.~Quintana-Ort{\'\i}, R.~van~de Geijn, F.~V. Zee,
  E.~Chan, Programming matrix algorithms-by-blocks for thread-level
  parallelism, {ACM} Trans. Mathematical Software 36~(3) (2009) 14:1--14:26.

\bibitem{BadiaHLPQQ09}
R.~M. Badia, J.~R. Herrero, J.~Labarta, J.~M. P\'erez, E.~S.
  Quintana-Ort\'{\i}, G.~Quintana-Ort\'{\i}, Parallelizing dense and banded
  linear algebra libraries using {SMPSs}, Concurrency and Computation: Practice
  and Experience 21 (2009) 2438--2456.

\bibitem{AliBBBQ14}
J.~I. Aliaga, R.~M. Badia, M.~Barreda, M.~Bollh\"ofer, E.~S.
  Quintana-Ort\'{\i}, Leveraging task-parallelism with {OmpSs} in {ILUPACK's}
  preconditioned {CG} method, in: 26th Int. Symp. on Computer Architecture and
  High Performance Computing (SBAC-PAD 2014), 2014, pp. 262--269.

\bibitem{agullo:hal-01333645}
E.~Agullo, A.~Buttari, A.~Guermouche, F.~Lopez,
  \href{http://doi.acm.org/10.1145/2898348}{Implementing multifrontal sparse
  solvers for multicore architectures with sequential task flow runtime
  systems}, ACM Trans. Math. Softw. 43~(2) (2016) 13:1--13:22.
\newblock \href {http://dx.doi.org/10.1145/2898348}
  {\path{doi:10.1145/2898348}}.
\newline\urlprefix\url{http://doi.acm.org/10.1145/2898348}

\bibitem{7965168}
J.~I. Aliaga, R.~Carratal\'a-S\'aez, R.~Kriemann, E.~S. Quintana-Ort\'{\i},
  Task-parallel {LU} factorization of hierarchical matrices using {OmpSs}, in:
  2017 IEEE International Parallel and Distributed Processing Symposium
  Workshops (IPDPSW), 2017, pp. 1148--1157.
\newblock \href {http://dx.doi.org/10.1109/IPDPSW.2017.124}
  {\path{doi:10.1109/IPDPSW.2017.124}}.

\bibitem{openmpweb}
The {OpenMP API} specification for parallel programming,
  \url{http://www.openmp.org/}.

\bibitem{ompssweb}
{OmpSs} project home page, \url{http://pm.bsc.es/ompss}.

\bibitem{7967171}
J.~M. Perez, V.~Beltran, J.~Labarta, E.~Ayguad\'e, Improving the integration of
  task nesting and dependencies in {OpenMP}, in: 2017 IEEE International
  Parallel and Distributed Processing Symposium (IPDPS), 2017, pp. 809--818.
\newblock \href {http://dx.doi.org/10.1109/IPDPS.2017.69}
  {\path{doi:10.1109/IPDPS.2017.69}}.

\bibitem{GVL3}
G.~Golub, C.~V. Loan, Matrix Computations, 3rd Edition, The Johns Hopkins
  University Press, Baltimore, 1996.

\bibitem{bem}
J.~T. Katsikadelis, Boundary Elements Theory and Applications, Elsevier, 2002.

\bibitem{H2Interpolation}
W.~Hackbusch, S.~Boerm,
  \href{http://www.sciencedirect.com/science/article/pii/S0168927402001216}{H2-matrix
  approximation of integral operators by interpolation}, Applied Numerical
  Mathematics 43~(1) (2002) 129 -- 143, 19th Dundee Biennial Conference on
  Numerical Analysis.
\newblock \href
  {http://dx.doi.org/https://doi.org/10.1016/S0168-9274(02)00121-6}
  {\path{doi:https://doi.org/10.1016/S0168-9274(02)00121-6}}.
\newline\urlprefix\url{http://www.sciencedirect.com/science/article/pii/S0168927402001216}

\bibitem{xeonplatinum}
A.~G\'omez-Iglesias, F.~Cheng, L.~Huan, H.~Liu, S.~Liu, C.~Rosales,
  Benchmarking the {Intel Xeon Platinum} 8160 processor, Tech. Rep. TR-17-01,
  Texas Advanced Computing Center, available at
  \url{https://repositories.lib.utexas.edu/handle/2152/61472} (2017).

\end{thebibliography}


%
%
%
\end{document}